\documentclass{aa}
\usepackage{graphicx}

\begin{document}

         
\title{The Globular Cluster System of \object{NGC\,1316} (Fornax A)}

\author{M. G\'omez\inst{1,2}
        \and
        T. Richtler\inst{3}
        \and
        L. Infante\inst{1}
        \and
        G. Drenkhahn\inst{4}}
 
      \offprints{mgomez@astro.puc.cl}
      
      \institute{Departamento de Astronom\'{\i}a y Astrof\'{\i}sica,
        P.~Universidad Cat\'olica de Chile. Casilla 306, Santiago,
        Chile\\
        \email{mgomez@astro.puc.cl,linfante@astro.puc.cl} \and
        Sternwarte der Universit\"at Bonn, Auf dem H\"ugel 71, D-53121
        Bonn, Germany \and Departamento de F\'{\i}sica, Universidad de
        Concepci\'on. Casilla 4009, Concepci\'on, Chile\\
        \email{tom@coma.cfm.udec.cl} \and Max-Planck-Institut f\"ur
        Astrophysik, Postfach 1317, D-85741 Garching bei M\"unchen,
        Germany\\
        \email{georg@mpa-garching.mpg.de}}


\date{Received ... / Accepted ...}

\abstract{ We have studied the Globular Cluster System of the merger
  galaxy \object{NGC\,1316} in Fornax, using CCD $BVI$ photometry.
  A clear bimodality is not detected from the broadband colours.
  However, dividing the sample into red (presumably metal-rich) and
  blue (metal-poor) subpopulations at $B-I=1.75$, we find that they
  follow strikingly different angular distributions. The red clusters
  show a strong correlation with the galaxy elongation, but the blue
  ones are circularly distributed. No systematic difference is seen in
  their radial profile and both are equally concentrated.
  We derive an astonishingly low Specific Frequency for
  \object{NGC\,1316} of only $S_{N}=0.9$, which confirms with a larger
  field a previous finding by Grillmair et~al. (\cite{grillmair99}).
  Assuming a ``normal'' $S_{N}$ of $\sim4$ for early-type galaxies, we
  use stellar population synthesis models to estimate in 2\,Gyr the
  age of this galaxy, if an intermediate-age population were to
  explain the low $S_{N}$ we observe. This value agrees with the
  luminosity-weighted mean age of \object{NGC\,1316} derived by
  Kuntschner \& Davies (\cite{kuntschner98}) and Mackie \& Fabbiano
  (\cite{mackie98}).
  By fitting $t_{5}$ functions to the Globular Cluster Luminosity
  Function (GCLF), we derived the following turnover magnitudes:
  $B=24.69\pm0.15$, $V=23.87\pm0.20$ and $I=22.72\pm0.14$. They
  confirm that \object{NGC\,1316}, in spite of its outlying location,
  is at the same distance as the core of the Fornax cluster.
%
}

\maketitle

\section{Introduction}

The analysis of globular cluster systems (GCSs) in elliptical galaxies
can have different motivations. One of them is to investigate the
variety of GCS morphologies in relation to their host galaxy
properties in order to gain insight into the formation of cluster
systems (see Ashman \& Zepf \cite{ashmanzepf97} and Harris
\cite{harris00} for reviews).

On the other hand, GCSs have been successfully employed as distance
indicators (Whitmore et~al. \cite{whitmore95a}, Harris
\cite{harris00}).  This is particularly interesting if the host galaxy
is simultaneously host for a type Ia supernova whose absolute
luminosity can accordingly be determined, as has been the case for
\object{SN\,1992A} in \object{NGC\,1380} (Della~Valle et~al.
\cite{dellavalle98}) and \object{SN\,1994D} in \object{NGC\,4526}
(Drenkhahn \& Richtler \cite{georg99}).

However, it can happen that both aspects are equally interesting as
with the target of the present contribution, \object{NGC\,1316}.

\object{NGC\,1316} (Fornax A), the brightest galaxy in the Fornax
cluster, is also one of the closest and brightest radio galaxies.
Catalogued as a S0 peculiar galaxy, it hosted two supernovae of type
Ia (\object{SN\,1980D} and \object{SN\,1981N}). The core of the Fornax
cluster is quite compact, which makes it a better extragalactic
distance standard than the Virgo cluster (see Richtler et~al.
\cite{tom99b}). However, \object{NGC\,1316} is located relatively far
out, $3\fdg7$ away from the central giant elliptical
\object{NGC\,1399}.

The early studies of Schweizer (\cite{schweizer80} \&
\cite{schweizer81}) indicated that \object{NGC\,1316} is very probably
a merger galaxy. Schweizer (\cite{schweizer80}) discovered the
presence of several irregular loops and tidal tails, as well as an
inclined disk of ionized gas rotating much faster than the stellar
spheroid. The age of the merger event was estimated to about $4\cdot
10^8$--$2\cdot10^9$ yrs.

Studies of galaxy mergers have shown (e.g. \object{NGC\,1275},
Holtzmann et~al. \cite{holtzman92}, \object{NGC\,7252}, Whitmore
et~al. \cite{whitmore93}, \object{NGC\,4038}/\object{NGC\,4039},
Whitmore \& Schweizer \cite{whitmore95b}) that bright massive star
clusters can form in large numbers during a merger event. But despite
the strong evidence for a previous merger in \object{NGC\,1316}, the
only indication for cluster formation is that Grillmair et~al.
(\cite{grillmair99}, hereafter Gr99), could not see a turnover in the
GCLF at the expected magnitude. They interpreted this finding as an
indication for an enhanced formation of many less massive clusters,
perhaps in connection with the merger. However, their HST study was
restricted to the innermost region of the galaxy.

In contrast to other galaxy mergers, where, presumably caused by the
high star-formation rate (Larsen \& Richtler \cite{soeren99},
\cite{soeren00}), the specific frequency of GCs increases,
Gr99\nocite{grillmair99} found an unusually \emph{small} total number
of clusters relative to the luminosity of \object{NGC\,1316}
($M_{V}\sim-22.8$, adopting a distance modulus to Fornax of
$\mu=31.35$, Richtler et~al. \cite{richtler00}).

There is also evidence from stellar population synthesis of integrated
spectra that \object{NGC\,1316} hosts younger populations (Kuntschner
\& Davies \cite{kuntschner98}, Kuntschner \cite{kuntschner00}) and
thus the question arises, whether the surprisingly low specific
frequency is caused by a high luminosity rather than by a small number
of clusters.  Goudfrooij et~al. (\cite{goud00}, hereafter Go00)
obtained spectra of 27 globular clusters and reported that the 3
brightest clusters have an age of about 3\,Gyr, indicating a high
star-formation activity 3\,Gyr ago, presumably caused by the merger
event.

These findings and the hope for a good distance via the GCLF were the
main motivation to do the present study of the GCS of
\object{NGC\,1316} in a larger area than that of the HST study. As we
will show, this galaxy resembles in many aspects the ``old'' merger
galaxy \object{NGC\,5018}, whose GCS has been investigated by Hilker
\& Kissler-Patig (\cite{hilker96}).

The paper is organized as follows: in Sect.~\ref{sec:obsred} we
discuss the observations, the reduction procedure and the selection of
cluster candidates.  The photometric and morphological properties of
the GCS are discussed in Sect.~\ref{sec:phot}.
Sect.~\ref{sec:specfreq} contains our findings concerning the Specific
Frequency.  We conclude this work with a general discussion in
Sect.~\ref{sec:disc}.

\section{Observations and Reduction}
\label{sec:obsred}

\begin{figure}
  \begin{center}
    \includegraphics[width=0.80\hsize]{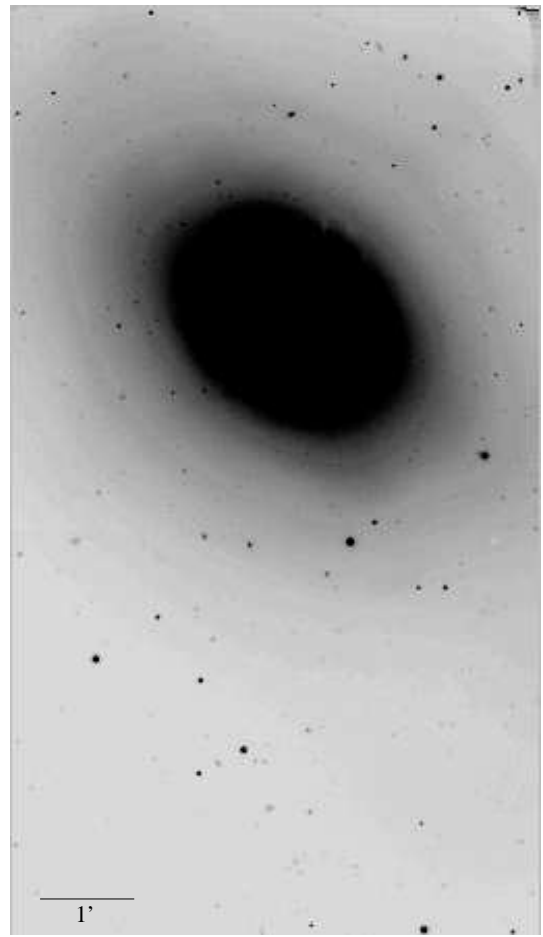}
  \end{center}
  \caption{The entire observed field in the $B$ filter, combined from both
    nights. North is up, east is left.}
  \label{matchedframes}
\end{figure}

The $B$, $V$ and $I$ images were obtained at the $3.6$\,m telescope at
La~Silla during the nights 29th and 30th of December, 1997 (dark
moon), using the ESO Faint Object Spectrograph and Camera, EFOSC2. The
field of view was $5\farcm6\times5\farcm6$ with a scale of
$0\farcs32$/pixel. During the first night, short- and long-exposures
in each filter were centred on the galaxy.  In the second night, a
background field located about $5\arcmin$ away from the centre of
\object{NGC\,1316} was observed, overlapping by $1\arcmin$ the
observations of the first night.  In addition, several fields
containing standard stars from the Landolt catalog (Landolt
\cite{landolt}) were acquired in each filter, as well as some short
exposures of \object{NGC\,1316}. Fig.~\ref{matchedframes} shows the
combined frames from both nights and Table~\ref{tab.observations}
summarises the observations.

\begin{table}
  \caption{Summary of the observations.}
  \label{tab.observations}
  \begin{center}
    \begin{tabular}{cccc}
      \hline
      \noalign{\smallskip}
      Date & Filter & Exposure (sec) & seeing \\
      \hline
      \noalign{\smallskip}
      Dec. 29, 1997 & $B$ & $4\times600$ & $1\farcs1$ \\
      Dec. 29, 1997 & $V$ & $5\times300$ & $1\farcs0$ \\
      Dec. 29, 1997 & $I$ & $6\times300$ & $1\farcs0$ \\
      Dec. 30, 1997 & $B$ & $4\times600$ & $1\farcs3$ \\
      Dec. 30, 1997 & $V$ & $3\times600$ & $1\farcs3$ \\
      Dec. 30, 1997 & $I$ & $3\times600$ & $1\farcs2$ \\
      \hline
    \end{tabular}
  \end{center}
\end{table}
 
\begin{table*}
  \caption{General parameters of the target galaxy, from
    de~Vaucouleurs et~al. (\cite{devaucouleur91}) and
    Poulain (\cite{poulain88}).}
  \label{tab.generalprop}
  \begin{center}
    \begin{tabular}{ccccccccc}
      \hline
      \noalign{\smallskip}
      name & $\alpha$ (2000) & $\delta$ (2000) & $l$ & $b$ & type & $m_V$ & $B-V$ & Veloc. [km/s]\\\hline
      \object{NGC\,1316} & $03^\mathrm{h} 22^\mathrm{m} 41\fs6$ & $-37\degr06\arcmin10\arcsec$ & $240\fdg16$ & $-56\fdg69$ & (R')SAB(s)0 & $8.53\pm0.08$ & 0.86  & $1793\pm12$ \\
      \hline
    \end{tabular}
  \end{center}
\end{table*}
 
\subsection{Reduction}

The frames were bias-subtracted and flat-fielded using standard IRAF
procedures. Bad pixels were replaced by interpolation with neighbour
pixels, using the task \emph{fixpix}. The frames in each filter were
then combined to remove the cosmic rays and to improve the S/N ratio.
Small shifts were applied when necessary to avoid misalignment.

To perform the photometry, a relatively flat sky background must be
provided. Therefore, the galaxy light needs to be subtracted.  While
in the case of early-type galaxies this is normally achieved by
fitting elliptical isophotes (see, e.g. Jedrzejewski
\cite{Jedrzejewski}), \object{NGC\,1316} exhibits some substructures
near its centre that can not be modeled in this way. We therefore
applied a median filter, using the task \emph{median} and a grid size
of $35\times35$\,pixels.

After the subtraction of each combined image from its corresponding
median frame, the object detection and PSF photometry was run using
DAOPHOT under IRAF.  The threshold level was set to $3\sigma$, and
typically 2500 to 3000 objects were detected in each image. Many of
these detections, particularly in the central parts of the galaxy, are
spurious because of the remaining photon noise after subtraction of
the galaxy model (see Sect.~\ref{sec:complcorr}). Since globular
clusters in the Fornax distance are not resolved, additional aperture
corrections are not necessary. The object lists of the three frames
were matched to remove spurious detections and to leave only objects
detected in all three filters. This reduced the sample to 911 objects.
We used the $(x,y)$ coordinates of this list as input for the program
SExtractor (Bertin \& Arnouts \cite{bertin96}), which performed the
classification of point-like and galaxy-like objects by means of a
``stellarity index'', calculated by a trained neural network from
ellipticity and sharpness.  Although SExtractor offers the possibility
of performing aperture photometry, we stress that we used it only to
get a better star-galaxy separation than is achievable with DAOPHOT
alone.

\subsection{Calibration of the photometry}

A large number of standard frames were acquired in each band during
the second night (the first night was not photometric). The colours of
the globular clusters are well inside the range spanned by the
standard stars. Aperture photometry with several annuli was performed
for every standard star to construct a growth-curve (see, e.g. Stetson
\cite{stetson90}). Only stars with well defined growth curves were
selected. From these curves we chose an aperture radius of 30 pixels
($9\farcs6$) for the photometry of the standard stars. We then fit the
instrumental magnitudes ($m_{j}$) to the equations:
\begin{eqnarray}
  \label{eq:photcal}
  m_{\rm b} & = & A_{\rm b0} + M_{\rm b} + A_{\rm b1}\cdot (B-V) + A_{\rm b2}\cdot X \\
  m_{\rm v} & = & A_{\rm v0} + M_{\rm v} + A_{\rm v1}\cdot (V-I) + A_{\rm v2}\cdot X \\
  m_{\rm i} & = & A_{\rm i0} + M_{\rm i} + A_{\rm i1}\cdot (V-I) + A_{\rm i2}\cdot X
\end{eqnarray}
where $A_{j0}$ and $M_{j}$ are the zeropoints and the standard
apparent magnitudes, respectively. $A_{j1}$ and $A_{j2}$ are the
colour and extinction coefficients. $X$ is the airmass.  The fitted
parameters are listed in Table~\ref{tab.photometry}.

\begin{table}
  \caption{Colour and extinction coefficients for our
    photometry according to (\ref{eq:photcal}).}
  \label{tab.photometry}
  \begin{center}
    \begin{tabular}{lccc}
      \hline
      \noalign{\smallskip}
      filter & $A_{j1}$ & $A_{j2}$ & rms of the fit \\
      \hline
      \noalign{\smallskip}
      $B$ & $-0.029\pm0.019$ & $0.220\pm0.038$ & $0.025$ \\
      $V$ & $-0.049\pm0.010$ & $0.111\pm0.017$ & $0.014$ \\
      $I$ & $~~~0.023\pm0.015$ & $0.046\pm0.029$ & $0.027$ \\
      \hline 
    \end{tabular}
  \end{center}
\end{table}

To check our photometry, we compared the magnitude and colours for
\object{NGC\,1316} quoted by Poulain (\cite{poulain88}) with our
values using the short-time exposures. The mean absolute differences
in magnitude and colour at five aperture radii between $22\farcs9$ and
$86\farcs6$ are: $\langle\Delta_{V}\rangle = 0.013$,
$\langle\Delta_{B-V}\rangle = 0.007$ and $\langle\Delta_{V-I}\rangle =
0.006$\,mag, well below the rms of the fit (see
Table~\ref{tab.photometry}).  We then defined 5 local standard stars
in the field of \object{NGC\,1316} to set the photometry of both
nights in a consistent way.

\subsection{Selection criteria}
\label{sec:selcrit}

Several criteria have been applied to select cluster candidates,
according to colours, magnitude, photometric errors, stellarity index
and projected position around the galaxy. We assume that the clusters
are similar to the Milky Way clusters.  Adopting an absolute turnover
magnitude (TOM) of $V=-7.60$ for the galactic clusters (Drenkhahn \&
Richtler \cite{georg99}, Ferrarese et~al.  \cite{ferrarese00}) and
$\mu = 31.35$ (Richtler et~al.  \cite{richtler00}), the TOM of
\object{NGC\,1316} is expected to be $V\sim23.7$\,mag, and the
brightest clusters about $V\sim20$.

\begin{figure}
  \includegraphics[width=\hsize]{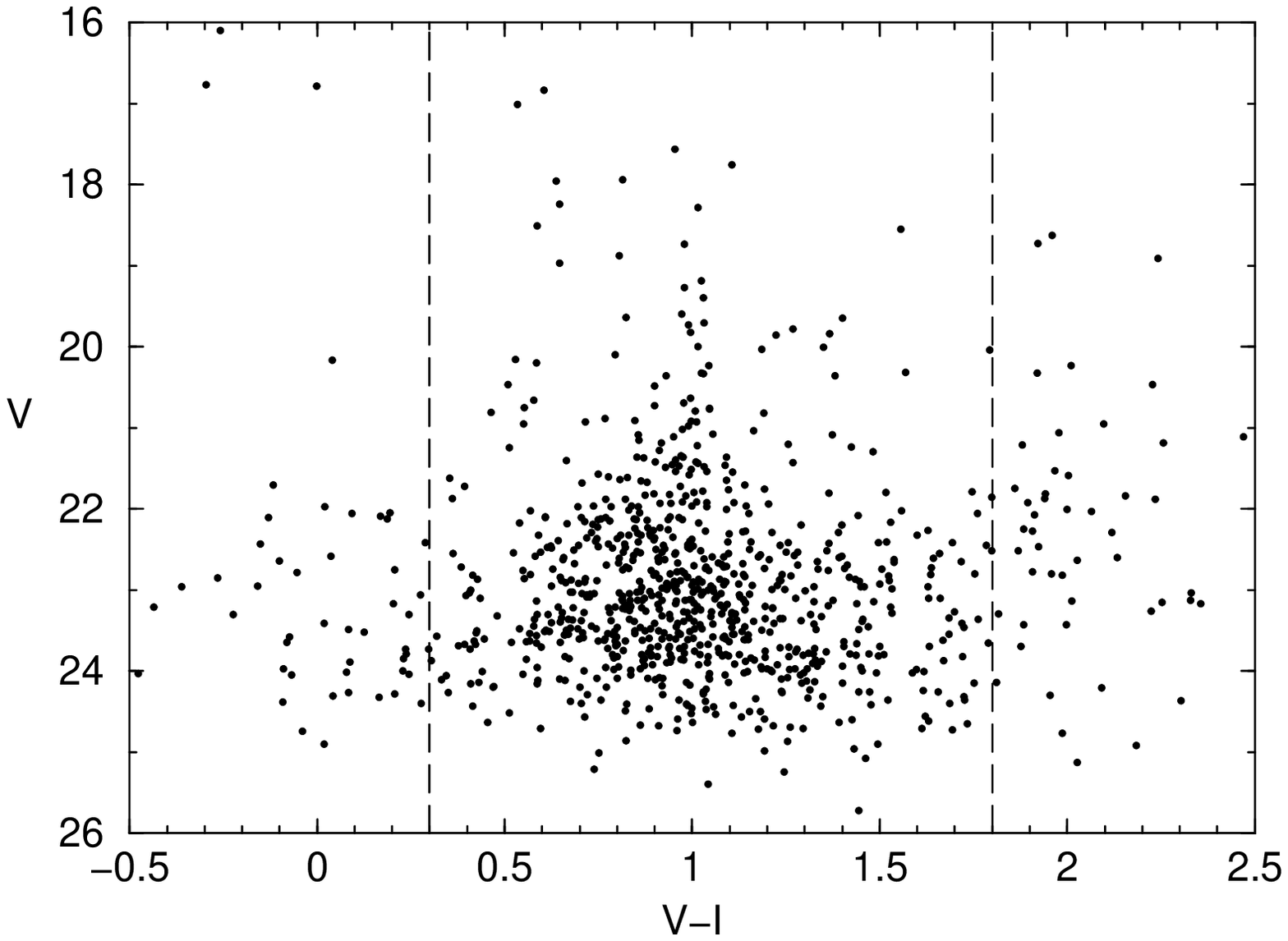}
  \caption{A colour-magnitude diagram for all objects detected in
    $B$, $V$ and $I$ in both nights before the selection criteria (911
    points)}. The dashed lines indicates the selection criterium in the
    $V-I$ colour (see Sect.~\ref{sec:selcrit}.)
  \label{colmagVI}
\end{figure}

Although no reddening corrections are normally applied when looking
towards Fornax (Burstein \& Heiles \cite{burstein82}), we cannot
restrict the colours of the clusters in \object{NGC\,1316} to match
exactly the galactic ones. One point is the smaller sample of
galactic clusters. Besides, we must allow for significant photometric
errors of faint cluster candidates.  Fig.~\ref{colmagVI} shows a
colour-magnitude diagram for all objects detected simultaneously in
$B$, $V$ and $I$ in both nights (before the selection), together with
the cut-off values adopted as criterium for this colour. As can be
seen, the majority of objects have colours around $V-I=1.0$.  Objects
bluer than 0.5\,mag are very probably foreground stars. A fraction of
the data points redder than $V-I=1.6$ are background galaxies.

We also tested in our images the robustness of the ``stellarity
index'' computed by SExtractor. This index ranges from $0.0$ (galaxy)
to $1.0$ (star) and varies for the same object by about $0.2$ when
classifying under different seeing conditions, except for the
brightest objects, which are clearly classified.  By visual inspection
of the images, we are quite confident that bright galaxies are always
given indices near to zero.  However, this classification becomes
progressively more difficult with fainter sources.

\begin{figure}
  \includegraphics[width=\hsize]{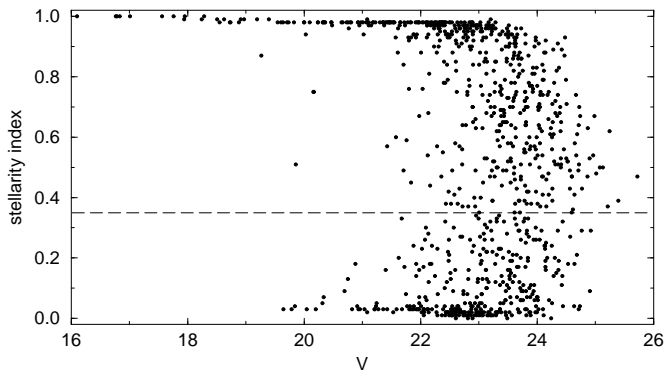}
  \caption{The stellarity index computed by SExtractor as function
    of the $V$ magnitude. The dashed lined indicates the cut value
    adopted in the selection criteria to reject background galaxies.}
  \label{stellarity}
\end{figure}

Fig.~\ref{stellarity} shows the ``stellarity index'' for the 911
objects in our sample (before the selection criteria) as a function of
the magnitude. Two groups of objects having indices of $\sim0.0$ and
$\sim1.0$ can be seen, but it is apparent that faint clusters cannot
be unambiguously distinguished from background galaxies due to the
uncertainty of the stellarity index in the case of faint sources.
Very similar results were obtained with our ``artificial stars'' (see
Sect.~\ref{sec:complcorr}), where indices down to 0.2 were measured
for faint objects that are constructed using the PSF model and,
therefore, are expected to have a stellarity index of $\sim 1.0$.
  
Guided by our experience with artificial stars, we defined the cut-off
value for the stellarity index to be 0.35. Remaining galaxies will be
statistically subtracted because the same criteria are applied to the
background field (see Sect.~\ref{sec:backcorr}).

Finally, we rejected objects with photometric error larger than
$0.15$\,mag.  To summarise, the following criteria were applied to our
sample of 911 sources detected in $B$, $V$ and $I$:
\newcounter{marker}
\begin{list}{\roman{marker})}{\usecounter{marker}}
\item $V > 19.5$
\item $0.4 < B-V < 1.4$
\item $0.3 < V-I < 1.8$
\item stellarity index $> 0.35$
\item error($V$), error($B-V$), error($V-I$) $< 0.15$
\end{list}
375 objects met this set of criteria and are our globular cluster
candidates.

\subsection{Background correction}
\label{sec:backcorr}
 
After the selection, there might still be some contamination by
foreground stars and background galaxies in our sample.  To subtract
them statistically, one needs to observe a nearby field, where no
clusters are expected, and to apply the same detection and selection
criteria as with the galaxy frame. However, the analysis of our
background field still shows a concentration towards the galaxy
centre, which means that only part of the field can be considered as
background.

We used the radial profile of the GC surface densities (see
Sect.~\ref{sec:radprof}) to select all objects on the flat part of the
profile, i.e., where the number of globular cluster candidates per
area unit exhibits no gradient. Fig.  \ref{radialprof} (top) shows
that this occurs at $r\approx300\arcsec$, where $r$ is the distance
from the optical centre.  Thus, all objects with galactocentric
distances larger than $300\arcsec$ and matching the above criteria,
constitute our background sample.

\begin{figure}
  \includegraphics[width=8.5cm]{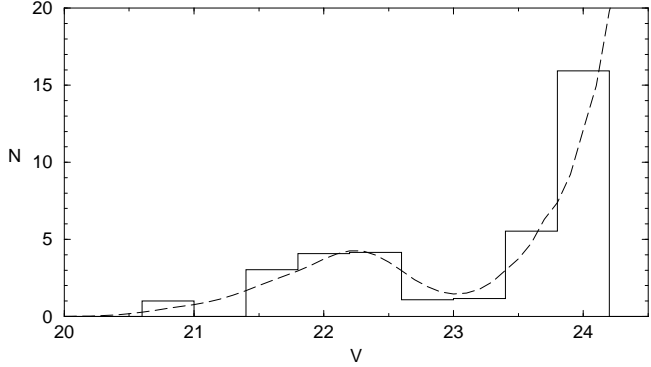}
  \caption{The semi-empirical luminosity function for the
    background field (dashed line). A histogram with a bin size of
    $0.4$\,mag is over-plotted for comparison.}
  \label{background}
\end{figure}

We constructed a semi-empirical luminosity function of the background
clusters with a technique described by Secker \& Harris
(\cite{secker93}), where a Gaussian is set over each data point,
centred at the corresponding observed magnitude. The sum of all
Gaussians gives a good representation of the background, without
introducing an artificial undulation or loss of information due to the
binning process. Fig.~\ref{background} shows the histogram of the
background objects, using a bin size of $0.4$\,mag, and the adopted
semi-empirical function, which we use as background in the calculation
of the luminosity function (see Sect.~\ref{sec:lumfunc}) and the
specific frequency. Admittedly, the numbers are small. The dip at
$V=23$ may be simply a result of bad statistics. On the other hand, as
Table 6 shows, the background counts are small compared to the
clusters counts. Therefore, errors of the order of the statistically
expected uncertainty do not significantly influence our results and
are accounted for in the uncertainties of the total counts in Table 6.

\subsection{Completeness correction}
\label{sec:complcorr}
 
To correct statistically for completeness, one needs to determine
which fraction of objects are actually detected by the photometry
routines. These `artificial stars experiments' were performed using
the task \emph{addstar} in DAOPHOT. In one step, 100 stars were added
in the science frames from $V=20$ to $V=25$ in steps of $0.1$\,mag,
distributed randomly to preserve the aspect of the image and not to
introduce crowding as an additional parameter, but with the same
coordinates in the $B$, $V$ and $I$ frames every loop.  The colours of
the stars were forced to be constant $B-V=0.7$ and $V-I=1.0$, close to
the mean colours of the globular clusters in \object{NGC\,1316} (see
Sect.~\ref{sec:coldist}). The object detection, photometry,
classification and selection criteria were applied in \emph{exactly
  the same way} as for the globular cluster candidates.  To get
sufficiently good statistics, we repeated the whole procedure 10
times, using different random positions in each of them and averaging
the results. In all, 50000 stars were added.

\begin{figure}
  \includegraphics[width=\hsize]{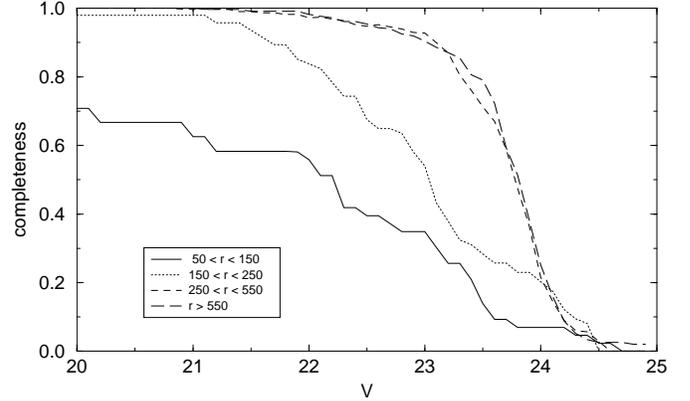}
  \caption{Completeness factors in four elliptical annuli. The
  probability of detection strongly decreases near the centre of 
  \object{NGC\,1316}.}
  \label{completeness}
\end{figure}

However, the completeness is not only a function of the magnitude.
Due to the remaining noise after galaxy subtraction, the probability
of detecting an object near the centre of the galaxy is smaller than
in the outer parts. We divided our sample of artificial stars into
elliptical rings, in the same way as we did in deriving the radial
profile (see Sect.~\ref{sec:radprof}). The results of the completeness
tests are summarised in Fig.~\ref{completeness}.  It can be seen that
the 50\% limit goes deeper with increasing galactocentric distance.

\section{Photometric and morphological properties}
\label{sec:phot}
 
\subsection{Colour distribution}
\label{sec:coldist}
 
In this section we discuss the colours of our GC candidates. No
interstellar reddening is assumed (Burstein \& Heiles
\cite{burstein82}) and therefore, only internal reddening might affect
the colours.

\begin{figure}
  \includegraphics[width=8.5cm]{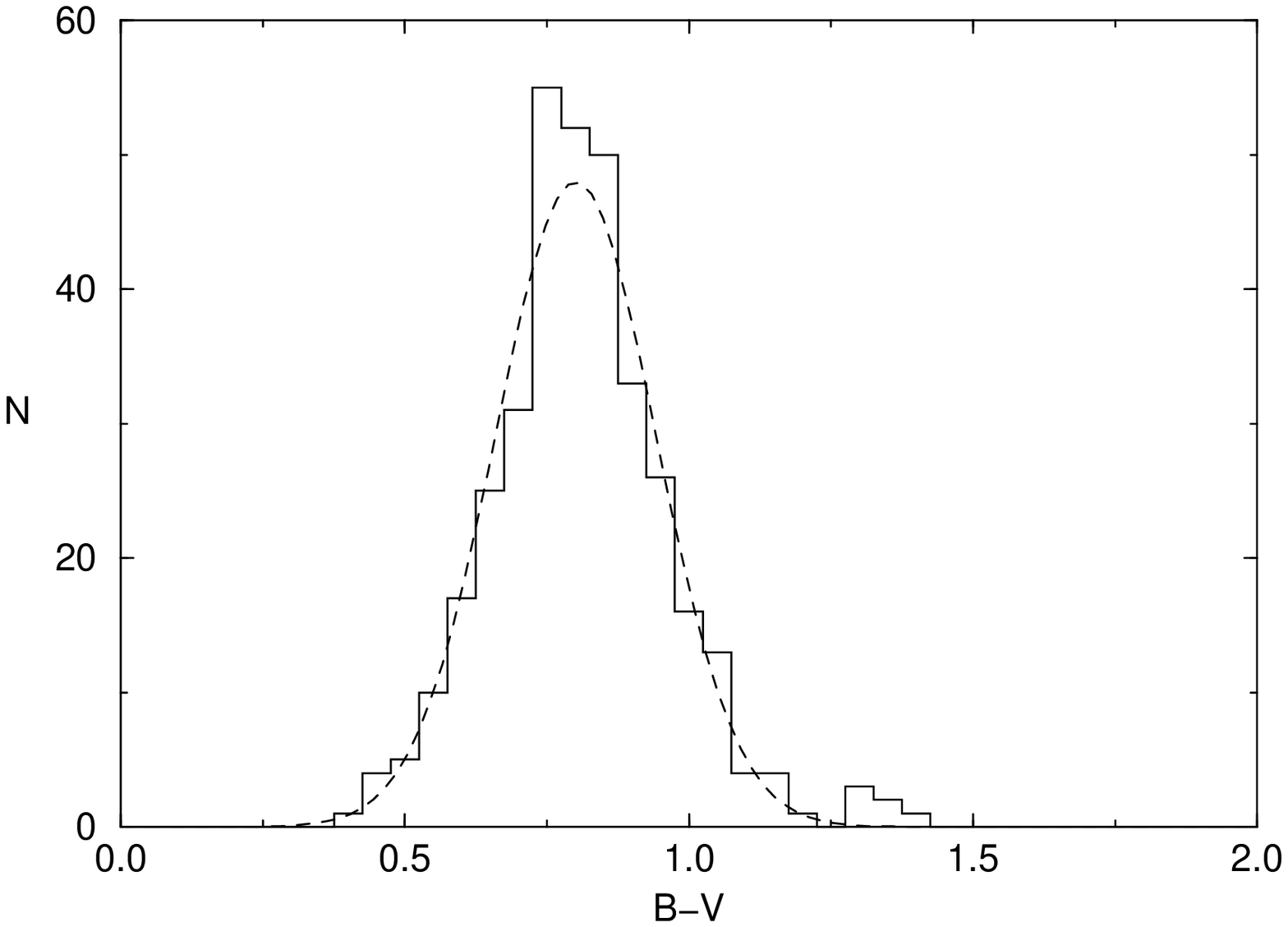}
  \includegraphics[width=8.5cm]{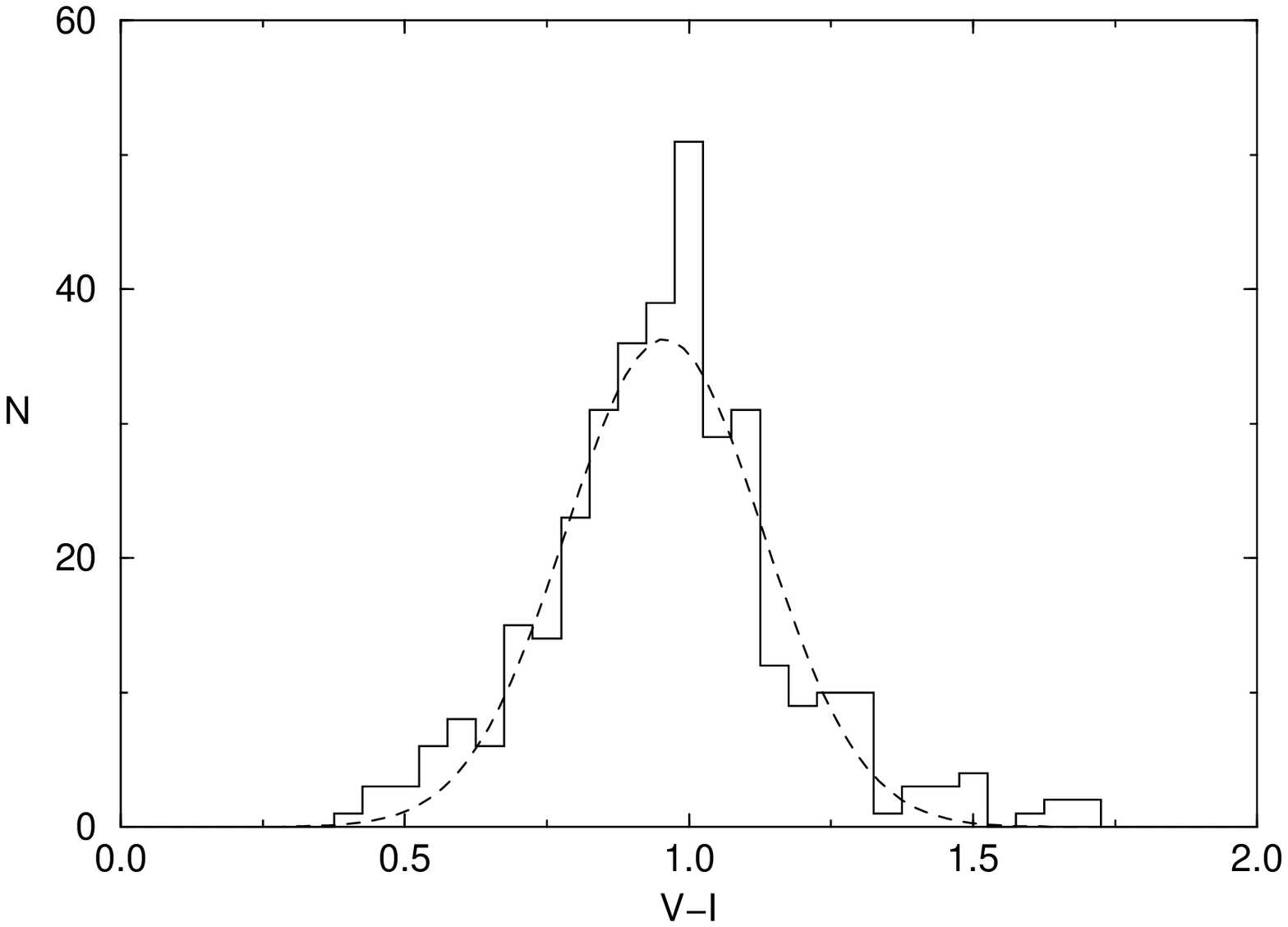}
  \includegraphics[width=8.5cm]{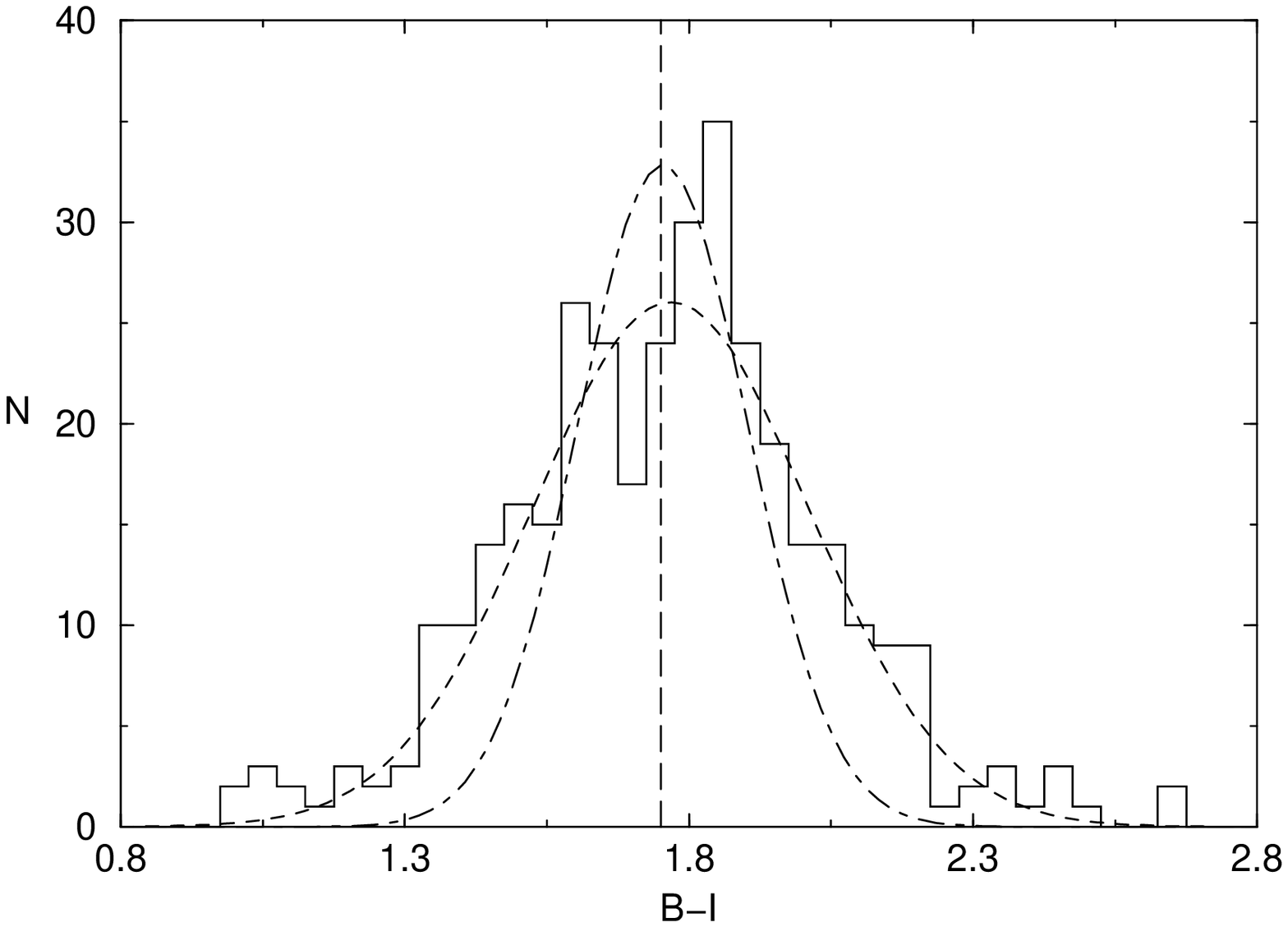}
  \caption{Colour histograms of the cluster candidates.
    The bin size is $0.05$\,mag in each colour. A Gaussian function
    has been fit to the histograms (dashed line, see text). The
    dot-dashed line is a Gaussian with $\sigma=0.15$. The long-dashed
    line at $B-I=1.75$ (lower panel) was set to divide the sample in
    red and blue clusters.}
  \label{colour_histograms}
\end{figure}

The histograms in Fig.~\ref{colour_histograms} show the colour
distribution of the cluster candidates.  A fit of a Gaussian function
with free dispersions returned the values listed in
Table~\ref{tab.colourgaussfit} and is overplotted for comparison.

\begin{table}
  \caption{Fit of a Gaussian function to the colour histograms
    of the GC candidates.}
  \label{tab.colourgaussfit}
  \begin{center}
    \begin{tabular}{lcc}
      \hline
      \noalign{\smallskip}
      colour & centre & $\sigma$ \\
      \hline
      \noalign{\smallskip}
      $B-V$ & $0.80\pm0.02$ & $0.13\pm0.04$ \\
      $V-I$ & $0.95\pm0.01$ & $0.17\pm0.02$ \\
      $B-I$ & $1.77\pm0.02$ & $0.35\pm0.03$ \\ 
      \hline 
    \end{tabular}
  \end{center}
\end{table}

\begin{figure}
  \includegraphics[width=\hsize]{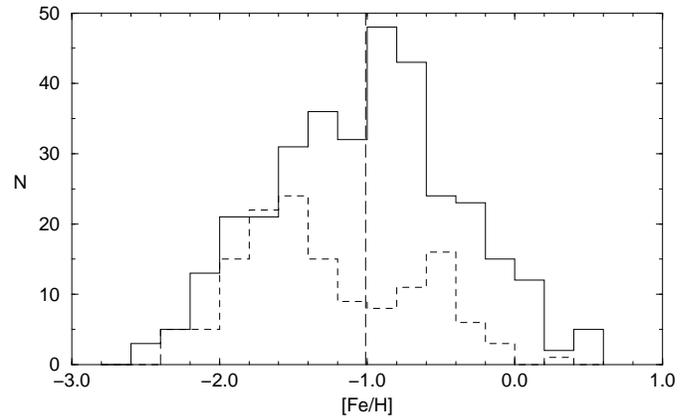}
  \caption{Metallicity histogram of the GCS of \object{NGC\,1316}, from
    its $B-I$ broadband colour. The galactic system (dashed line) is
    overplotted for comparison. The long-dashed line near
    $[\element{Fe}/\element{H}]=-1.0$ divides the sample in metal-rich
    and metal-poor clusters, using the calibration from Couture et~al
    (\cite{couture90}) and $B-I=1.75$ (see text and
    Fig.~\ref{colour_histograms}.)}
  \label{metal_histo}
\end{figure}

As can be seen, the dispersion in the $B-V$ and $V-I$ histograms is
completely explained by the photometric errors alone. This is not the
case for the $B-I$ histogram, due probably to the greater metallicity
sensitivity of this colour. A Gaussian with $\sigma=0.15$\,mag (the
maximum allowed photometric error, see Sect.~\ref{sec:selcrit}) is
overplotted and is clearly narrower than the colour distribution of
the GCs.  We use the $B-I$ broadband colour to derive a mean
metallicity, as it offers the largest baseline and minimizes the
effect of random photometric errors. From the calibration of Couture
et~al.  (\cite{couture90}) for globular clusters in the Milky Way:
\begin{equation}
  \label{eq:bi}
  (B-I)_0 = 0.375\cdot[\element{Fe}/\element{H}] + 2.147
\end{equation}
the mean derived metallicity is $[\element{Fe}/\element{H}] =
-1.01\pm0.05$\,dex, where the uncertainty reflects only the error in
the fit of a Gaussian function to the colour histogram. The
corresponding metallicity histogram is shown in
Fig.~\ref{metal_histo}. The clearly bimodal distribution of the
galactic clusters is overplotted for comparison (dashed line). We
cannot exclude that the distribution is bimodal, but then the
metal-poor and the metal-rich peak are closer to each other than in
the case of the galactic system.

\begin{figure}
  \includegraphics[width=\hsize]{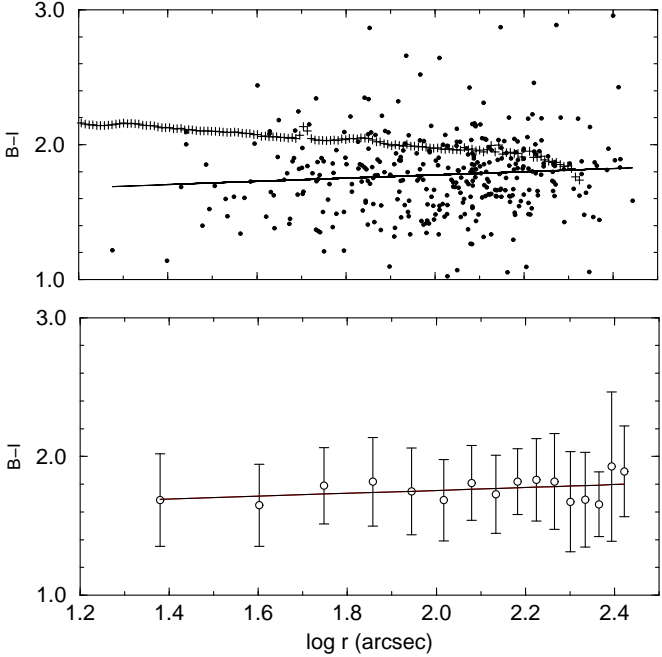}
  \caption{The $B-I$ colour gradient along the projected
    galactic radius. Top: the cluster candidates and a least-square
    fit(solid line). The crosses indicate the $B-I$ colour of the
    galaxy. Bottom: the mean colour in several rings, with the error
    bars indicating the $\sigma$ of the mean at that ring.}
  \label{colourgradient}
\end{figure}

To search for colour gradients, we plotted (Fig.~\ref{colourgradient})
the $B-I$ colour for the candidates, discarding the zone of the
background population. Although a least-square fit to a function of
the form $B-I = A\cdot 10^{m\cdot r}$gives $\Delta(B-I)/\Delta\log (r)
= 0.12 \pm 0.07$, the data scatter too much to conclude definitely
that such a gradient exists. A similar result was obtained by dividing
the sample of clusters in circular rings and calculating the mean
$B-I$ colour for each ring. The slope of $\Delta(B-I)/\Delta\log (r) =
0.11 \pm 0.07$ agrees very good with the previous fit, but it is
strongly determined by the first and the last two points, where the
statistic is poor.  From the same plot, it is apparent that the
clusters are on average $\sim0.2$\,mag bluer than the underlying
galaxy light. This is a property commonly found also in normal
early-type galaxies, where the blue and presumably metal-poor clusters
indicate the existence of a faint metal-poor stellar (halo?)
population, as the case of \object{NGC\,1380} suggests (Kissler-Patig
et.~al \cite{kissler97}). However, the non-existence of a colour
gradient is consistent with the finding that blue and red clusters
have similar surface density profiles (Sect.~\ref{sec:radprof}).

Differential reddening caused by the irregular dust structure might
affect the width of the colour distribution, but apparently does not
produce any colour gradient.

\subsection{Angular distribution}
\label{sec:angulardist}

For all objects in our sample of clusters, a transformation from
cartesian $(x,y)$ coordinates to polar $(r,\theta)$ was done, with
origin in the optical centre of \object{NGC\,1316}. An offset of
$50\degr$ in $\theta$ was applied to match the PA of the galaxy quoted
by RC3. In this way, $\theta = 0$ represents the direction of the
semi-major axis (sma) of the galaxy, $a$, and $\theta=90\degr$ the
direction of $b$, the semi-minor axis.

To analyse the angular distribution, we rejected objects inside a
radius of 150 pixels (corresponding to 4.3\,kpc with $\mu=31.35$),
where the completeness is significantly lower (see
Fig.~\ref{completeness}) and some clusters appear over ripples and
dust structures. Objects outside of 450 pixels were also rejected, as
one needs equally-sized sectors to do this analysis, and $r=450$ is
roughly the radius of the largest circle fully covered by our frame
(see dotted line in Fig.~\ref{isofotas}).  Only candidates brighter
than $V=23.8$ (the 50\% completeness level in the entire frame) were
considered.

\begin{figure}
  \includegraphics[width=\hsize]{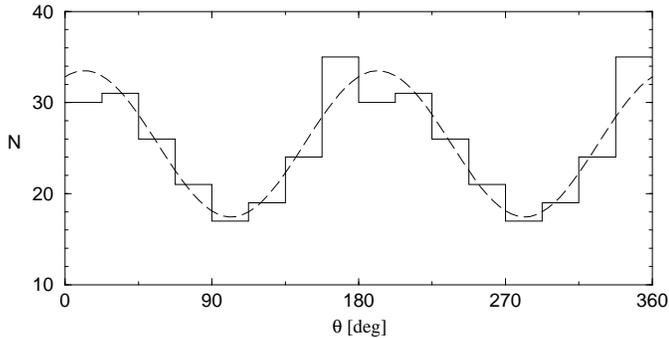}
  \caption{The angular distribution of globular clusters for
    a radius from 150 to 450 pixels ($48\arcsec$ to $144\arcsec$) down
    to $V=23.8$.  The bin size is $22\fdg5$ and the data were taken
    mod~$\pi$. The histogram from $0\degr$ to $180\degr$ is repeated
    from $180\degr$ to $360\degr$ for a better visualisation.  The
    dashed line indicates the best fit of a double-cosine function.}
  \label{angular}
\end{figure}

The data were then binned in $\theta$. Several bin sizes from
$18\degr$ to $30\degr$ were tested, and Fig.~\ref{angular} shows the
histogram for a bin size of $22\fdg5$, which corresponds to dividing
the sample into 16 sectors. The bins were taken modulo $\pi$ for a
better statistics, that is, we assume that the distribution of
clusters is symmetric along the semi-mayor axis of the galaxy. Due to
the substructures present in \object{NGC\,1316}, and from the fact
that it is a merger galaxy, one could expect some systematic
differences in the azimuthal distribution of the clusters between both
halves.  Although this is not observed at any bin size, the small
number of counts does not allow us to address this question clearly.
 
The histogram shown in Fig.~\ref{angular} demonstrates the strong
correlation of the globular clusters with the galaxy light. By fitting
isophotes to the galaxy, we obtained ellipticities ranging from $0.27$
to $0.32$ and position angles from $49\degr$ to $53\degr$, between
semi-major axes of 150 to 450 pixels (the same used with the
clusters).

From the least-square fit to this histogram (with fixed period $\pi$),
we derived $\mathrm{PA}=63\degr\pm9\degr$ and an ellipticity of
$0.38\pm0.06$.
 
There are, however, some problems that cannot be easily resolved with
our ground-based data. As indicated, the detection of cluster
candidates near the centre of the galaxy is quite poor. This is,
unfortunately, the most interesting region to search for young
clusters which might be related to a merger event.

\subsection{Radial profile}
\label{sec:radprof}
 
\begin{figure}
  \begin{center}
    \includegraphics[width=0.8\hsize]{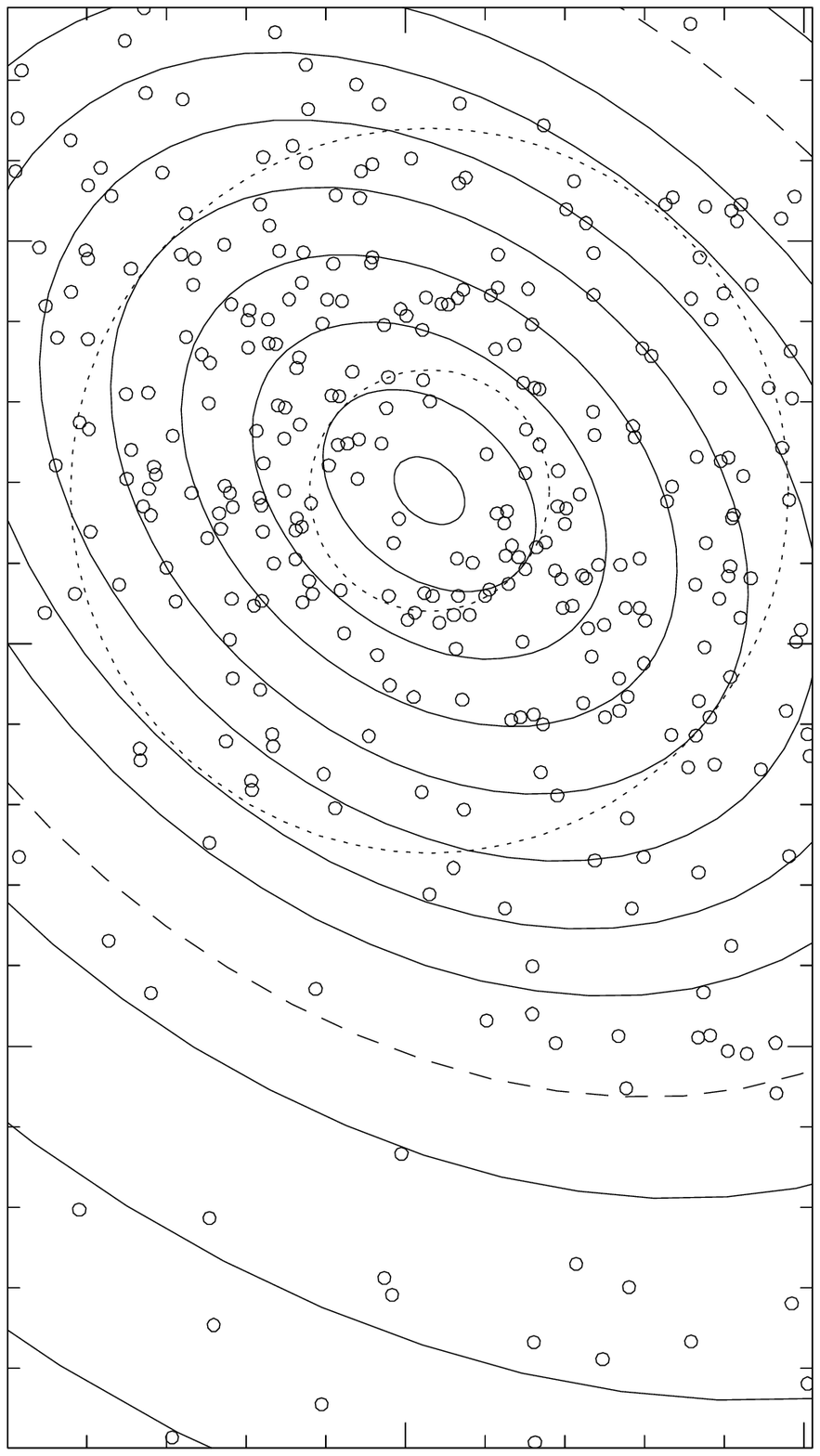}
  \end{center}
  \caption{The elliptical annuli used in the calculation of
    the globular cluster density and specific frequency. The open
    circles indicate the position of all GC candidates brighter than
    $V=23.8$. Objects outside the annulus defined by $r=150$ and
    $r=450$ pixels (dotted line) were rejected for the analysis of the
    angular distribution (see Sect.~\ref{sec:angulardist}). Objects
    outside the dashed ellipse are considered as background. The
    scale and orientation is the same as in the
    Fig.~\ref{matchedframes}.}
  \label{isofotas}
\end{figure}

To derive the radial surface density of GCs, we divide our sample into
elliptical annuli, as shown in Fig.~\ref{isofotas}.  The annuli have a
width of 100 pixels along the major axis ($a$) and start from $a=50$
pixels.  Due to the small number of objects in the periphery of
\object{NGC\,1316}, rings beyond $a=900$ were given a width of 300
pixels.

The ellipticity, position angle and centre of these rings were taken
from the fit of the galaxy light (see Sect.~\ref{sec:angulardist}) and
fixed for all annuli. In particular, the ellipticity $e$, defined as
$1-(b/a)$, where $a$ and $b$ are the semi-major and semi-minor axis
respectively, was set to 0.3 and the PA to $50\degr$.
 
We then counted the number of globular cluster candidates per unit
area in each ring, down to $V=23.8$, and corrected them with the
corresponding completeness function.
 
\begin{figure}
  \includegraphics[width=\hsize]{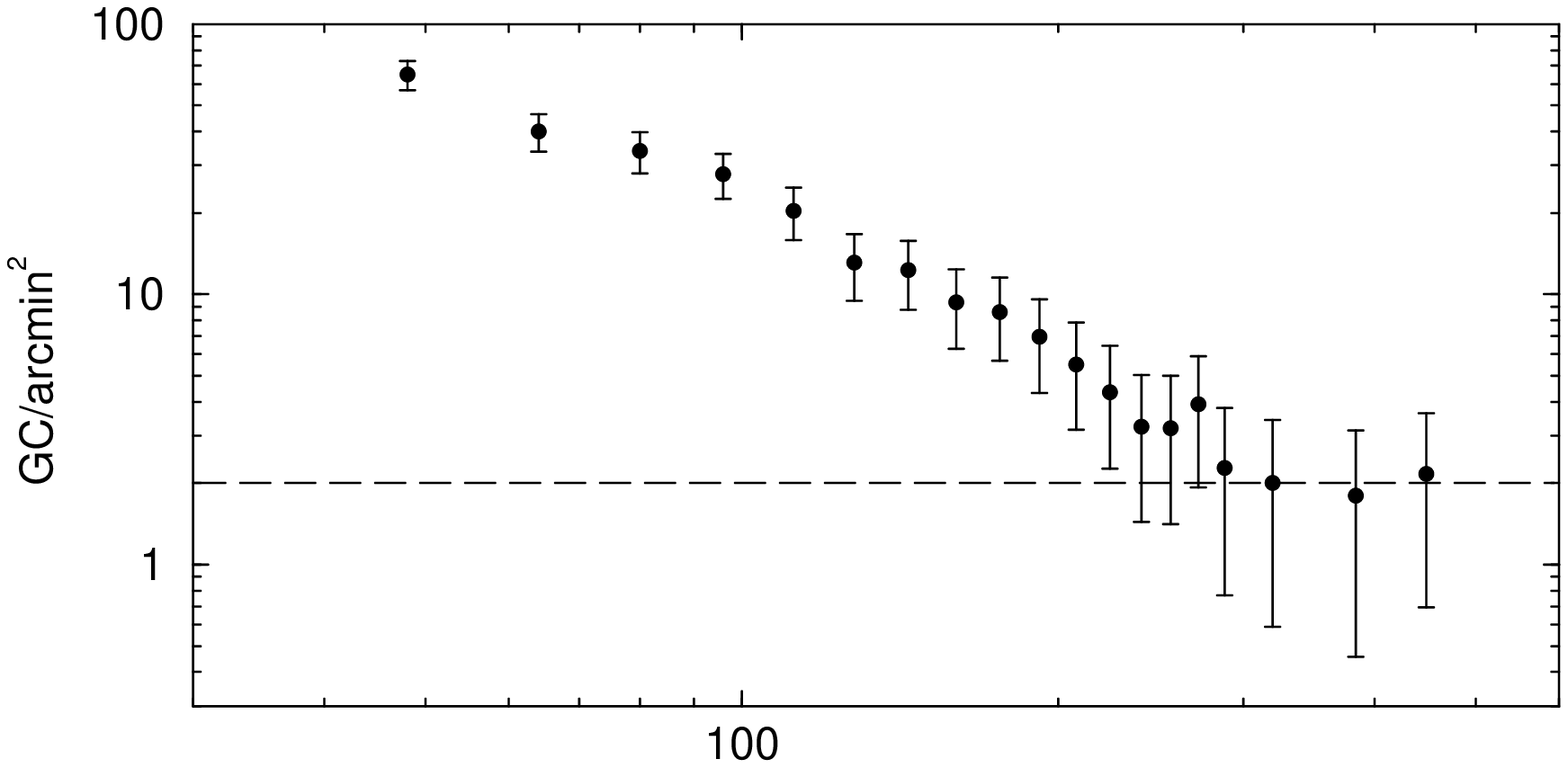}
  \includegraphics[width=\hsize]{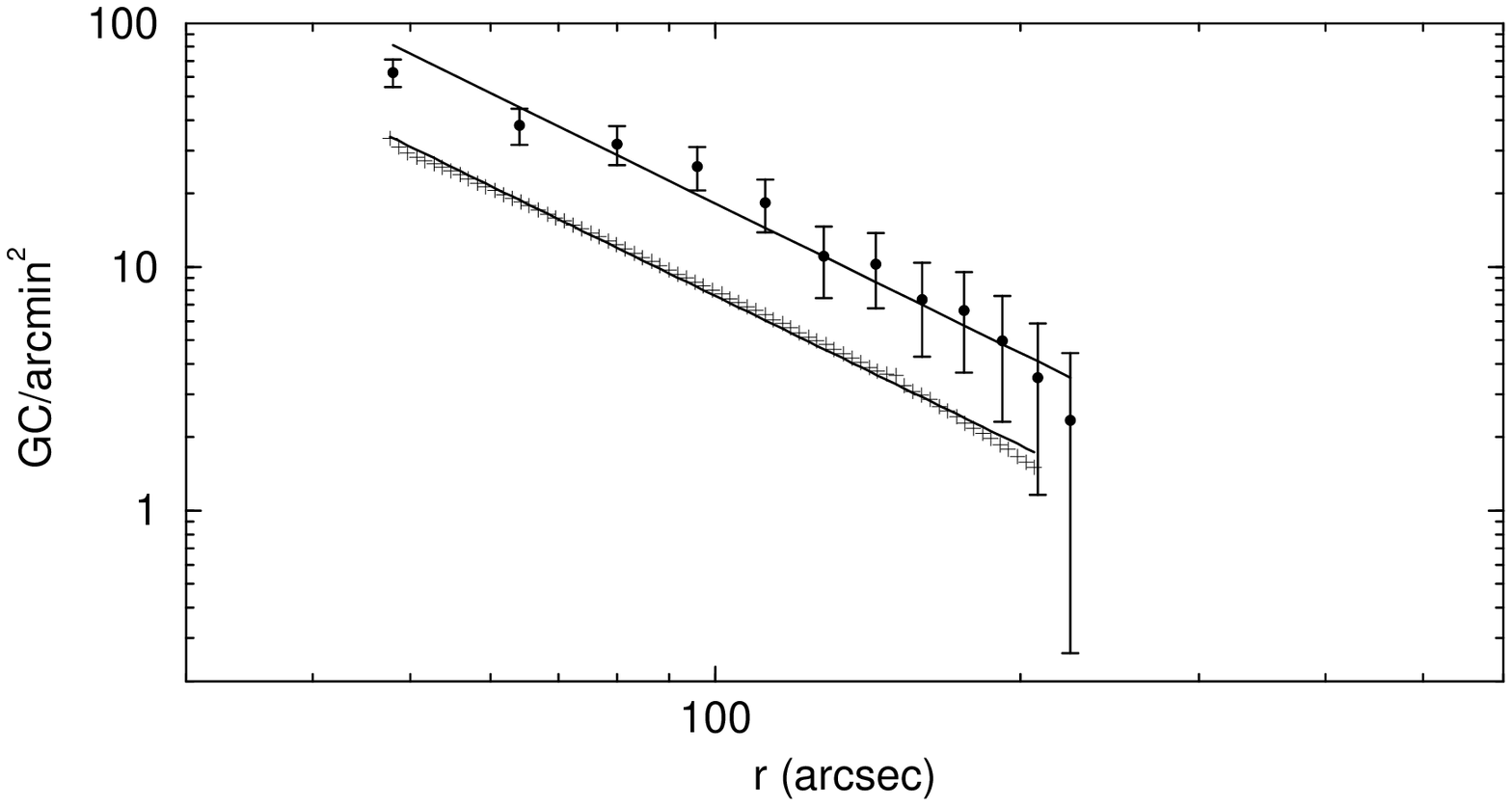}
  \caption{Top: the radial profile of the density of globular
    cluster candidates. The dashed line indicates the value adopted
    for the background, corresponding to $2.0$\,objects$/\sq\arcmin$.
    Bottom: the radial profile after subtraction of the background
    counts.  The crosses represent the profile of the galaxy light,
    arbitrarily shifted.}
  \label{radialprof}
\end{figure}

The results are presented in Table~\ref{tab.radialprof}. The first
column lists the mean semi-major axis of the ring. Column~2 gives the
raw number of clusters down to $V=23.8$, without correction for
completeness. Column~3 lists the corrected data with their errors.
Column~4 lists the visible area of the rings, in $\sq\arcmin$.
Column~5 gives the number of candidates per unit area.  Column~6, 7
and 8 are used to compute the Specific Frequency (see
Sect.~\ref{sec:specfreq}).  Fig.~\ref{radialprof} shows the radial
profile of the clusters' surface density, before and after the
subtraction of the background counts.

\begin{table*}
  \caption{This table gives the result of the radial profile of 
    the cluster candidate surface density. The first column lists the
    centre of each annulus (in pixels).  Then follows the raw number
    counts down to $V=23.8$, before and after the correction for
    completeness. The fourth column gives the visible area of the
    annulus in $\sq\arcmin$. Column~5 lists the mean density of GC per
    $\sq\arcmin$. Columns~6 and 7 give the number of clusters down to
    the TOM in $V$, before and after the correction for
    completeness. The applied geometrical corrections are listed in
    column 8. Finally, the number of the clusters in each annulus,
    after doubling the counts around the TOM. Note that the corrected
    and total number of clusters for the innermost annulus (sma=100)
    are NOT derived using the completeness correction. Instead, they
    were estimated by extrapolating the radial profile towards the
    centre (see text in Sect.~\ref{sec:specfreq}).}
  \label{tab.radialprof}
  \begin{center}
    \begin{tabular}{lcrcrcrcl}
      \noalign{\smallskip}
      \hline
      \noalign{\smallskip}
      sma & $N^\mathrm{raw}_{23.8}$ & $N^\mathrm{corr}_{23.8}$
      & area of ring & GC/area
      & $N^\mathrm{raw}_\mathrm{TOM}$ & $N^\mathrm{corr}_\mathrm{TOM}$
      & geom. factor & $N_\mathrm{annulus}$ \\
      {}[pixels] & & & [$\sq\arcmin$] & [$1/\sq\arcmin$] & & & & \\
      \noalign{\smallskip}
      \hline
      \noalign{\smallskip}
      100 & 20 & $58.6\pm18.1$  & 1.251 & $46.86\pm14.47$ & 20  & $200.2\pm37.5$ & 1.    & $400\pm75$ \\
      200 & 53 & $100.3\pm17.9$ & 2.502 & $40.09\pm7.15$  & 53  & $100.7\pm18.0$ & 1.    & $201\pm36$ \\
      300 & 68 & $104.4\pm15.2$ & 3.751 & $27.84\pm4.05$  & 69  & $109.4\pm15.8$ & 1.    & $219\pm32$ \\
      400 & 54 & $65.4\pm10.7$  & 5.006 & $13.06\pm2.13$  & 56  & $68.9\pm11.0$  & 1.    & $134\pm22$ \\
      500 & 48 & $58.2\pm10.0$  & 6.234 & $9.34\pm1.60$   & 51  & $62.0\pm10.4$  & 0.997 & $124\pm21$ \\
      600 & 37 & $42.9\pm8.5$   & 6.159 & $6.96\pm1.22$   & 37  & $44.5\pm8.8$   & 0.821 & $108\pm21$ \\
      700 & 19 & $22.4\pm6.2$   & 5.145 & $4.35\pm1.21$   & 19  & $22.8\pm6.3$   & 0.588 & $78\pm21$ \\
      900 & 19 & $23.2\pm6.4$   & 9.440 & $2.45\pm0.68$   & --- & ---            & ---   & --- \\
      1200& 11 & $12.7\pm4.6$   & 7.159 & $1.77\pm0.64$   & --- & ---            & ---   & --- \\
      1500& 4  & $5.3\pm3.2$    & 2.880 & $1.84\pm1.11$   & --- & ---            & ---   & --- \\
      \noalign{\smallskip}
      \hline
    \end{tabular}
  \end{center}
\end{table*}
 
A fit of a power-law $\rho(r) = A\cdot r^\alpha$, where $\rho$ is the
surface density and $r$ the projected distance along the semi-major
axis, gives $\alpha_\mathrm{gcs}=-2.04\pm0.20$ and
$\alpha_\mathrm{gal}=-2.03\pm0.02$ for the clusters and the galaxy
light, respectively.  We are well aware that a King profile may be
more adequate than a power function, but our purpose is to compare our
result with previous work in other GCSs in Fornax, which quote only
power functions.
 
The similarity between both slopes indicates that the GCS of
\object{NGC\,1316} is not more extended than the galaxy light.
Moreover, $\alpha_\mathrm{GCS}=-2.04$ is in good agreement with other
GCS of ``normal'' early-type galaxies in Fornax (Kissler-Patig et~al.
\cite{markus97}).

\begin{figure}
  \includegraphics[width=\hsize]{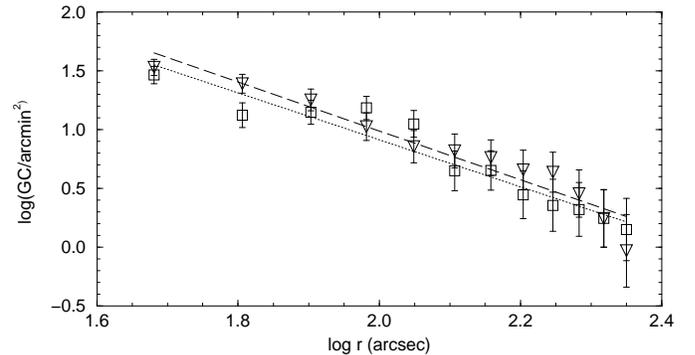}
  \caption{The radial profile of the GC surface density for the
    red (triangles) and blue population (squares).  No systematic
    difference is seen and both are equally concentrated.}
  \label{radialprof_rb}
\end{figure}

We divided the system at $B-I=1.75$\,mag into blue (presumably
metal-poor) and red (metal-rich) clusters, and searched for systematic
differences in the morphological properties between both subgroups.
Again, only clusters brighter than $V=23.8$ were considered, leaving
150 blue and 166 red clusters.  Our results show that both
subpopulations are equally concentrated towards the centre and share
the same density profile (see Fig.~\ref{radialprof_rb}).

\begin{figure}
  \includegraphics[width=\hsize]{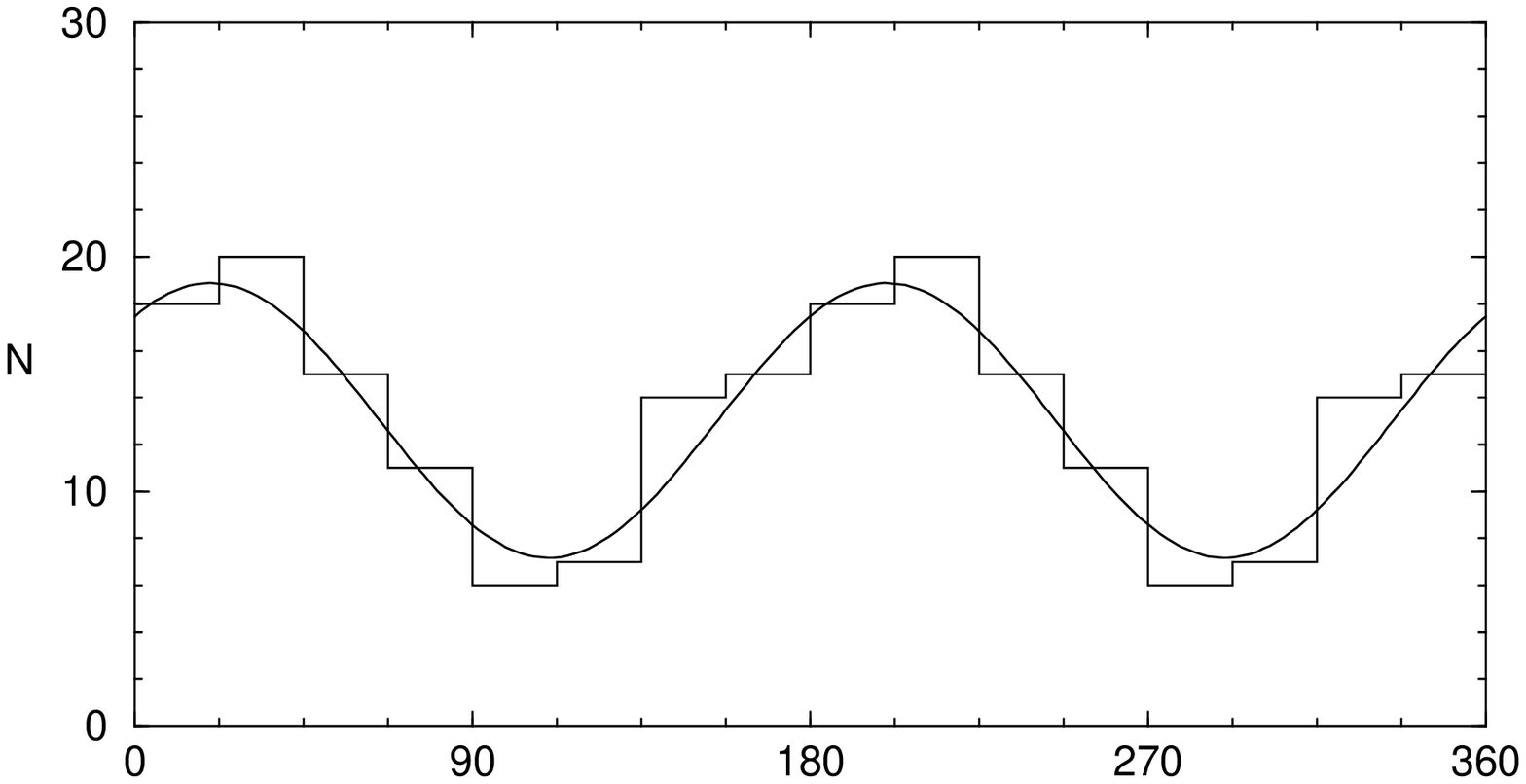}
  \includegraphics[width=\hsize]{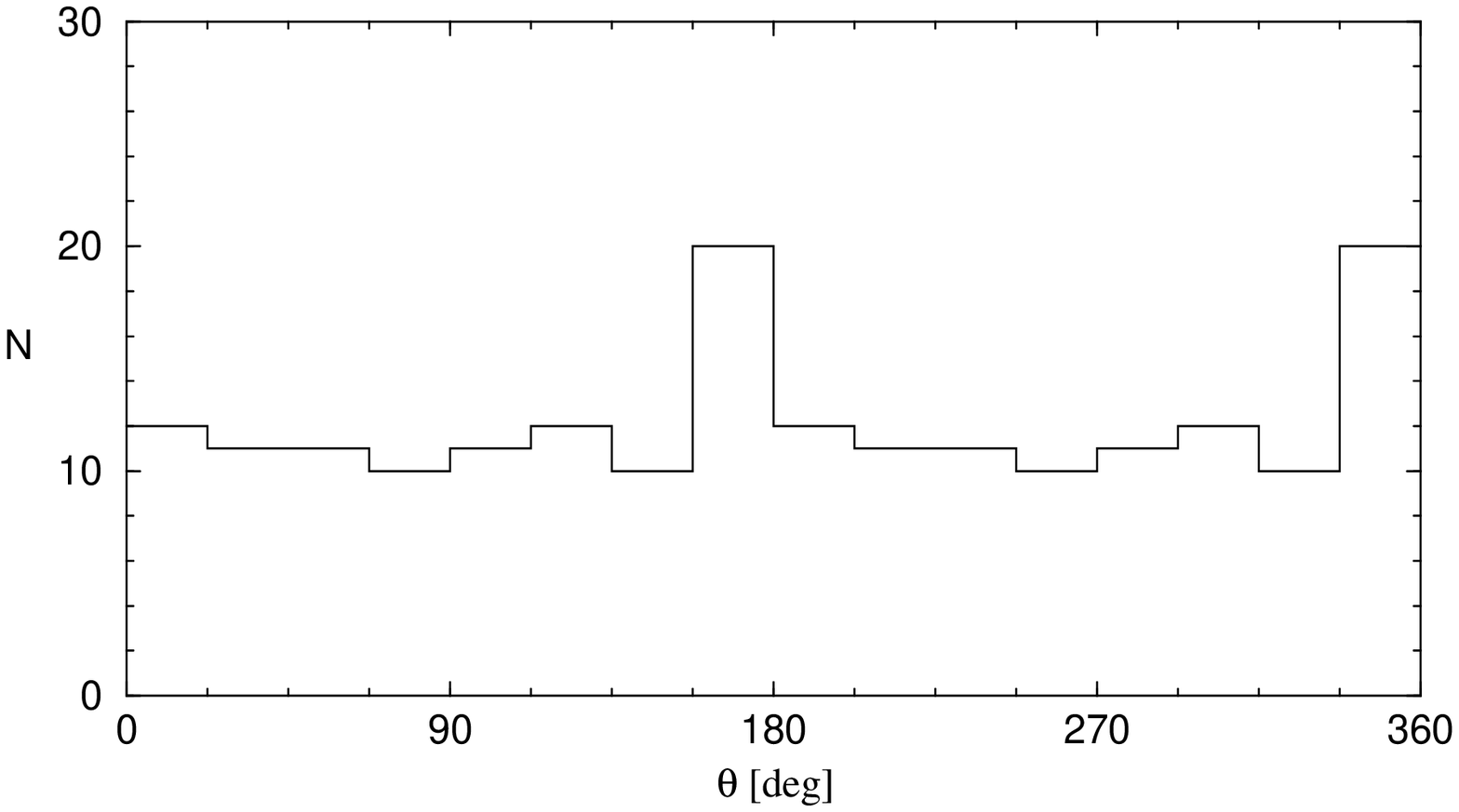}
  \caption{The azimuthal distribution of the red (top) and blue
    (bottom) clusters.  The bin size is $22\fdg5$. As with
    Fig.~\ref{angular}, the data are repeated from $180\degr$ to
    $360\degr$ for a better visualisation. A fit of a double-sin
    function is overplotted for the red sample.}
  \label{angular_rb}
\end{figure}

In contrast to the radial profile, red and blue clusters
systematically differ in their azimuthal distribution, as shown in
Fig.~\ref{angular_rb}. The redder clusters show a clear correlation
with the galaxy elongation, but the bluer appear to be spherically
distributed, resembling a halo and bulge population.  Similar results
were obtained at several bin sizes.
 
\subsection{The luminosity function (LF)}
\label{sec:lumfunc}

The GCLF has been widely used as a distance indicator.  The
theoretical reasoning for an universality of the TOM requires an
initially universal mass function of GCs in combination with a
subsequent universal dynamical evolution of the individual clusters of
a GCS (Okazaki \& Tosa \cite{okazaki95}, hereafter OT95, see e.g.
Ashman \& Zepf \cite{ashmanzepf97} and Harris \cite{harris00} for
reviews). While the latter is plausible, as far as the average
evolution is concerned, the former still lacks a theoretical
explanation.  Nevertheless, the empirical evidence for an universal
mass function as a power law $\mathrm{d}N/\mathrm{d}M \sim M^{-1.8}$,
where $\mathrm{d}N$ is the number of clusters in the mass range
$M+\mathrm{d}M$, is considerable.  GCSs of host galaxies of very
diverse nature such as the Milky Way, ellipticals in the Virgo cluster
(Harris \& Pudritz \cite{harris94}, McLaughlin \& Pudritz
\cite{mclaughlin96}) or young clusters in merging systems like
\object{NGC\,4038}/\object{NGC\,4039} (the Antenna, Whitmore \&
Schweizer \cite{whitmore95b}) or \object{NGC\,7252} (Miller et~al.
\cite{miller97}), all show mass functions which are well described by
the above power law.

However, the TOM must be a function of time, being fainter for younger
cluster systems (OT95\nocite{okazaki95}). This is due to the dynamical
dissolution of clusters, which affects first the less massive ones and
then progresses towards higher masses.  If, for example, the red
population or a part of it was formed in the merger, one would
accordingly expect a fainter TOM.  Therefore it is reasonable to give
the LF separately for the total system, for the red, and the blue
subpopulation.

The counts are given in Table~\ref{tab.lf}.  As stated in
Sect.~\ref{sec:backcorr}, only objects closer to the centre than
$r=300\arcsec$ were considered possible members of the GCS of
\object{NGC\,1316}. Again, $r$ is measured along the major axis of an
ellipse with ellipticity 0.3. The objects with $r>300\arcsec$ define
the background sample.  The background area is $13.72\sq\arcmin$,
while the area of the GCS is $38.01\sq\arcmin$.  We then used the
results of our artificial star experiments to derive the completeness
factors in four elliptical annuli: $16\arcsec<r<48\arcsec$,
$48\arcsec<r<80\arcsec$, $80\arcsec<r<176\arcsec$ and
$176\arcsec<r<300\arcsec$.  As can be seen from
Fig.~\ref{completeness}, the completeness corrections differ at least
in the three inner regions.  The bin centers are given in column~1.
Columns~2, 4, 6 and 8 list the counts in each region as explained
above, while columns~3, 5, 7, 9 contain the corresponding completeness
factors. The background counts, as defined in
Sect.~\ref{sec:backcorr}, are given in column~10.  The total number of
clusters is listed in column~11. This is the sum of the clusters in
each region (corrected for completeness), minus the background level
scaled to the area of counting.  The uncertainties result from an
error propagation, where also the error in the background was
considered. A value of 0.02 has been adopted as the error in the
completeness factors, which is roughly the RMS we observed between the
tests.

There are two additional columns, namely the number of blue and red
clusters. They were calculated in the same way and will be discussed
in Sect.~\ref{sec:disc}.

Bin sizes of $0.3$, $0.4$ and $0.5$\,mag were tested, and the results
were always in good agreement with each other.  Finally, we have
chosen $0.5$\,mag as our bin size from the appearance of the fit
histogram, and the absence of undulations which are present for the
other cases.

We note, however, that our derived TOM is close to the limit of the
observations, and the last bins are strongly affected by the
completeness correction.  Nevertheless, the similarity of the results
using different bin sizes and centers, and the robustness of the fit
against skipping the last bin, is encouraging.

\begin{table*}
  \caption{The counts (in bins of $0.5$\,mag) used in the
    determination of the GCLF. Given are the $V$-magnitudes (or $B$,
    $I$, respectively) of the bin centers. Then follow the raw
    counts of the four elliptical annuli $N_i$ (see text) together with
    the corresponding completeness factors $f_i$. The background counts
    are then listed as defined in Sect.~\ref{sec:backcorr}. The
    completeness factors of the fourth annulus have been used for
    the background as well. The total number of clusters is the sum
    of the four annuli minus the background, normalised to the same
    area (see text).  Also given is the number of clusters for the
    blue and the red population separately, where the separating
    colour was $B-I=1.75$.}
  \begin{center}
    \begin{tabular}{ccccccccccccc}
      \hline
      \noalign{\smallskip}
      $V$ & $N_1^V$ & $f_1$ & $N_2^V$ & $f_2$ & $N_3^V$ & $f_3$ &
      $N_4^V$ & $f_4$ & $N_\mathrm{bkg.}^V$ & $N_\mathrm{total}^V$ & $N_\mathrm{blue}^V$ & $N_\mathrm{red}^V$ \\
      \hline
      \noalign{\smallskip}
      19.5 & 0 & 0.71 & 1  & 0.98 & 1  & 1.00 & 2  & 1.00 & 0.0 & $4.0\pm2.0$    & $1.0\pm1.0$   & $3.0\pm1.7$ \\
      20.0 & 1 & 0.71 & 1  & 0.98 & 3  & 1.00 & 1  & 1.00 & 0.0 & $6.4\pm2.7$    & $2.4\pm1.7$   & $4.0\pm2.0$ \\
      20.5 & 1 & 0.67 & 0  & 0.98 & 4  & 1.00 & 2  & 1.00 & 0.2 & $6.9\pm3.1$    & $3.9\pm2.6$   & $3.0\pm1.7$ \\
      21.0 & 2 & 0.62 & 6  & 0.98 & 3  & 1.00 & 3  & 1.00 & 1.0 & $12.6\pm5.0$   & $1.4\pm3.3$   & $11.2\pm3.8$ \\
      21.5 & 2 & 0.58 & 3  & 0.94 & 12 & 0.99 & 7  & 0.99 & 2.5 & $18.8\pm7.0$   & $7.8\pm4.6$   & $11.3\pm5.2$ \\
      22.0 & 2 & 0.56 & 5  & 0.84 & 14 & 0.97 & 11 & 0.98 & 4.6 & $22.2\pm8.8$   & $14.4\pm6.9$  & $8.0\pm5.4$ \\
      22.5 & 7 & 0.40 & 11 & 0.68 & 32 & 0.95 & 14 & 0.95 & 4.2 & $69.8\pm12.5$  & $36.6\pm9.0$  & $33.3\pm8.6$ \\
      23.0 & 3 & 0.35 & 18 & 0.54 & 39 & 0.93 & 14 & 0.90 & 1.7 & $94.2\pm12.9$  & $44.3\pm8.7$  & $49.9\pm9.5$ \\
      23.5 & 2 & 0.14 & 8  & 0.28 & 52 & 0.71 & 19 & 0.79 & 3.7 & $127.2\pm19.9$ & $65.5\pm15.6$ & $61.7\pm12.1$ \\
      24.0 & 0 & 0.07 & 2  & 0.20 & 32 & 0.22 & 8  & 0.25 & 3.8 & $145.1\pm38.8$ & $48.8\pm24.8$ & $96.5\pm28.1$ \\
      24.5 & 0 & 0.00 & 0  & 0.00 & 4  & 0.03 & 1  & 0.03 & 0.5 & ---            & ---           & --- \\
      \hline
    \end{tabular}
    \label{tab.lf}
  \end{center}

  \begin{center}
    \begin{tabular}{ccccccccccccc}
      \hline
      \noalign{\smallskip}
      $B$ & $N_1^B$ & $f_1$ & $N_2^B$ & $f_2$ & $N_3^B$ & $f_3$ &
      $N_4^B$ & $f_4$ & $N_\mathrm{bkg.}^B$ & $N_\mathrm{total}^B$ & $N_\mathrm{blue}^B$ & $N_\mathrm{red}^B$\\
      \hline
      \noalign{\smallskip}
      20.5 & 0 & 0.71 & 1  & 0.98 & 2  & 1.00 & 2  & 1.00 & 0.0 & $5.0\pm2.2$     & $1.0\pm1.0$   & $4.0\pm2.0$ \\
      21.0 & 2 & 0.67 & 1  & 0.98 & 3  & 1.00 & 3  & 1.00 & 0.2 & $9.5\pm3.6$     & $5.4\pm3.1$   & $4.0\pm2.0$ \\
      21.5 & 1 & 0.67 & 3  & 0.98 & 4  & 1.00 & 1  & 1.00 & 0.7 & $7.6\pm4.0$     & $0.1\pm2.7$   & $7.6\pm2.9$ \\
      22.0 & 2 & 0.58 & 4  & 0.96 & 5  & 1.00 & 4  & 1.00 & 0.7 & $14.7\pm5.0$    & $7.1\pm3.9$   & $7.6\pm3.1$ \\
      22.5 & 2 & 0.58 & 5  & 0.89 & 15 & 0.98 & 8  & 0.99 & 3.3 & $23.2\pm7.9$    & $10.5\pm7.0$  & $12.8\pm5.3$ \\
      23.0 & 5 & 0.42 & 9  & 0.74 & 24 & 0.96 & 11 & 0.96 & 4.8 & $46.7\pm11.1$   & $32.0\pm9.7$  & $13.8\pm6.5$ \\
      23.5 & 4 & 0.35 & 11 & 0.64 & 31 & 0.94 & 17 & 0.92 & 3.2 & $70.4\pm12.0$   & $35.2\pm8.6$  & $35.3\pm9.1$ \\
      24.0 & 3 & 0.26 & 14 & 0.32 & 50 & 0.80 & 15 & 0.85 & 2.3 & $127.9\pm17.8$  & $57.8\pm16.5$ & $70.1\pm13.1$ \\
      24.5 & 1 & 0.07 & 6  & 0.23 & 47 & 0.47 & 13 & 0.51 & 4.8 & $139.8\pm27.6$  & $66.7\pm33.2$ & $73.1\pm16.4$ \\
      25.0 & 0 & 0.05 & 1  & 0.09 & 15 & 0.06 & 7  & 0.05 & 2.0 & $290.3\pm159.0$ & ---           & --- \\
      25.5 & 0 & 0.00 & 0  & 0.00 & 0  & 0.00 & 1  & 0.02 & 0.1 & ---             & ---           & --- \\
      \hline
    \end{tabular}
  \end{center}

  \begin{center}
    \begin{tabular}{ccccccccccccc}
      \hline
      \noalign{\smallskip}
      $I$ & $N_1^I$ & $f_1$ & $N_2^I$ & $f_2$ & $N_3^I$ & $f_3$ &
      $N_4^I$ & $f_4$ & $N_\mathrm{bkg.}^I$ & $N_\mathrm{total}^I$ & $N_\mathrm{blue}^I$ & $N_\mathrm{red}^I$\\
      \hline
      \noalign{\smallskip}
      18.5 & 0 & 0.71 & 1  & 0.98 & 0  & 1.00 & 2  & 1.00 & 0.0 & $3.0\pm1.7$    & $0.0\pm0.0$    & $3.0\pm1.7$ \\
      19.0 & 0 & 0.71 & 1  & 0.98 & 4  & 1.00 & 1  & 1.00 & 0.0 & $6.0\pm2.5$    & $1.0\pm1.0$    & $5.0\pm2.2$ \\
      19.5 & 3 & 0.67 & 1  & 0.98 & 3  & 1.00 & 0  & 1.00 & 0.0 & $8.5\pm3.3$    & $4.0\pm2.3$    & $4.5\pm2.3$ \\
      20.0 & 0 & 0.63 & 5  & 0.98 & 3  & 1.00 & 5  & 1.00 & 0.5 & $11.7\pm4.1$   & $3.2\pm2.5$    & $8.5\pm3.3$ \\
      20.5 & 4 & 0.58 & 4  & 0.94 & 11 & 0.99 & 5  & 0.99 & 2.8 & $19.8\pm7.4$   & $9.5\pm4.9$    & $10.3\pm5.5$ \\
      21.0 & 2 & 0.56 & 6  & 0.84 & 16 & 0.97 & 10 & 0.98 & 3.8 & $26.7\pm8.5$   & $9.3\pm5.4$    & $17.4\pm6.6$ \\ 
      21.5 & 4 & 0.40 & 9  & 0.68 & 35 & 0.95 & 17 & 0.95 & 5.1 & $63.1\pm12.1$  & $33.0\pm9.0$   & $30.1\pm8.0$ \\
      22.0 & 5 & 0.35 & 18 & 0.54 & 37 & 0.93 & 12 & 0.90 & 1.9 & $94.9\pm13.5$  & $39.0\pm9.2$   & $55.9\pm9.8$ \\
      22.5 & 1 & 0.14 & 9  & 0.28 & 46 & 0.71 & 15 & 0.79 & 2.3 & $115.0\pm17.9$ & $62.0\pm13.9$  & $53.4\pm10.9$ \\
      23.0 & 1 & 0.07 & 0  & 0.20 & 38 & 0.22 & 15 & 0.25 & 4.3 & $199.4\pm45.4$ & $130.6\pm36.4$ & $68.8\pm24.6$ \\
      23.5 & 0 & 0.02 & 1  & 0.00 & 3  & 0.03 & 0  & 0.03 & 1.4 & ---            & ---            & ---\\
      \hline
    \end{tabular}
  \end{center}
\end{table*}

\begin{figure}
  \includegraphics[width=\hsize]{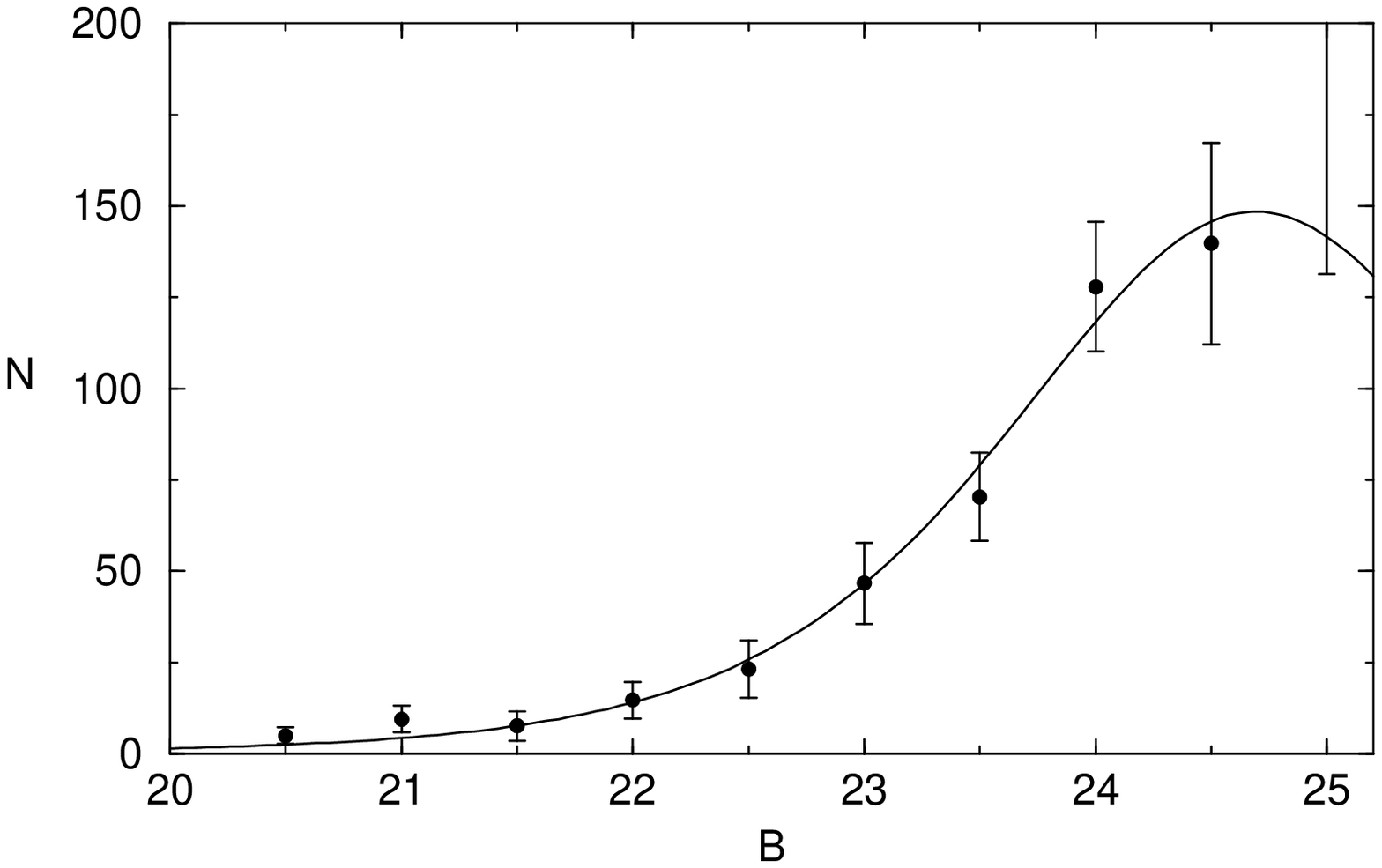}
  \includegraphics[width=\hsize]{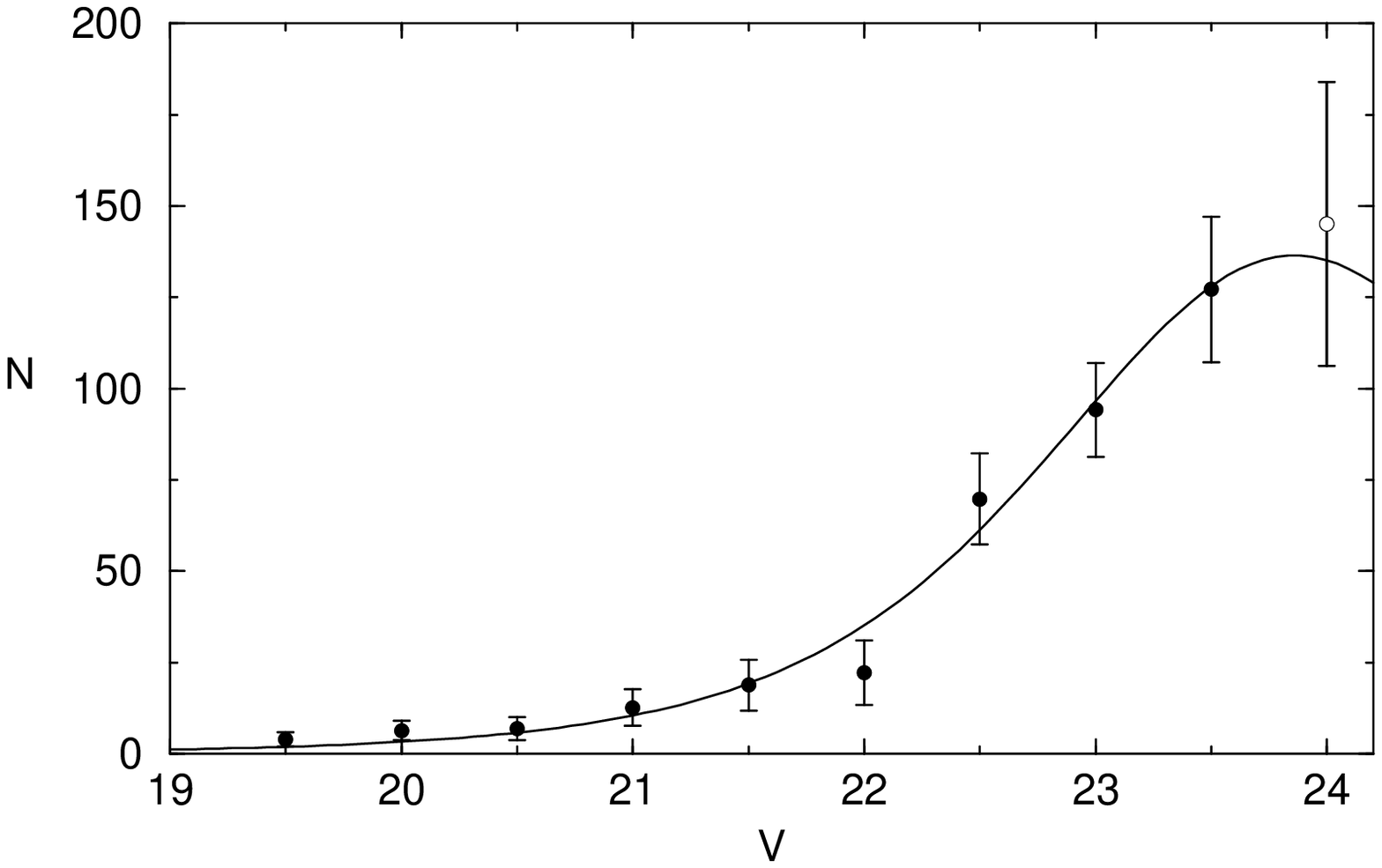}
  \includegraphics[width=\hsize]{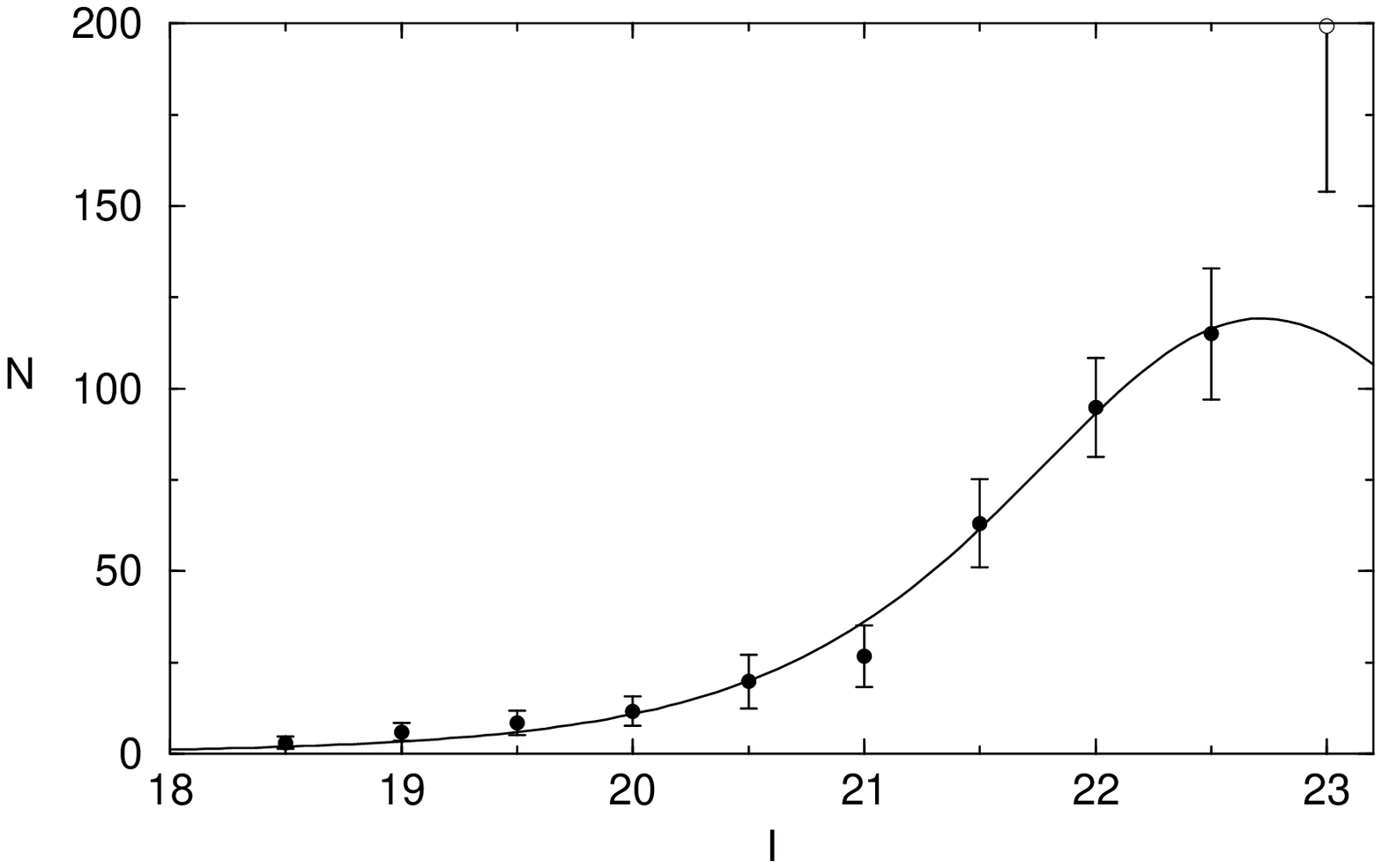}
  \caption{The luminosity functions for $B$, $V$ and $I$.
    The bin size is $0.5$\,mag. The last bin (open circle) was not
    used in the fit, as its center is beyond the limiting magnitude.
    The smooth curve is the fit of a $t_{5}$ function with
    $\sigma=1.1$.}
  \label{lumfun}
\end{figure}

Fig.~\ref{lumfun} shows the three luminosity functions in $B$, $V$,
$I$. For fitting the LF we chose $t_5$ functions, which are of the
form:
\begin{equation}
  \label{eq:t5}
  t_{5}(m) = \frac{8}{3\sqrt{5}\pi\sigma_{t}}
  \left(1+\frac{(m-m_0)^{2}}{5\sigma_{t}^{2}} \right)^{-3}
\end{equation}

Gaussians give no significantly different results, as earlier
investigations have demonstrated (e.g. Della~Valle et~al.
\cite{dellavalle98}). Since our photometry is not deep enough to fit
$\sigma_{t}$ independently, we used a fixed value of $\sigma_{t}=1.1$,
which is appropriate for early-type galaxies (Harris \cite{harris00}
quotes $1.4$ as a typical Gaussian dispersion, while $1.29$ is the
ratio between the Gaussian dispersion and the dispersion of a
$t_5$-function).

Table~\ref{tab.toms} lists the TOMs for the three bands $B$, $V$ and
$I$, additionally the TOMs for the blue and the red population in all
filters.  The uncertainties are the formal fit errors accounting also
for the count uncertainties in the individual bins, as given in
Table~\ref{tab.lf}.

It is apparent that the TOMs are not convincingly reached in the full
sample.  Moreover, it is clear that we cannot say anything definitive
about a possibly different TOM for the red and the blue population.
Deeper photometry is required.

The formal distance moduli, resulting from the full sample, are $31.58
\pm 0.18$ in $B$, $31.47 \pm 0.22$ in $V$, and $31.19 \pm 0.17$ in
$I$, adopting as the absolute TOMs $M^{T}_{B} = -6.89 \pm 0.10$,
$M^{T}_{V} = -7.60 \pm 0.08$ and $M^{T}_{I} = -8.47 \pm 0.10$
(Drenkhahn \& Richtler \cite{georg99}, Ferrarese et~al.
\cite{ferrarese00}).  Table~8 compiles the presently available TOMs of
other Fornax galaxies, published by several groups. The comparison
shows that the $B$ and the $V$ TOM match the average values of the
literature, while the $I$ value falls at the faint end. However, our
photometry is not sufficiently deep, and we are not able to assign a
strong independent weight to \object{NGC\,1316}.  Leaving open the
problem with the deviating $I$ TOM, we can only conclude that this
galaxy has the same distance as the other Fornax galaxies.  Since a
discussion of the distance scale uncertainty is beyond the scope of
our paper, we adopt in the following a distance modulus of 31.35\,mag
(Richtler et~al. \cite{richtler00}).

The contrasting finding of Gr99\nocite{grillmair99}, who were unable
to constrain the TOM in spite of their deeper data, may indicate that
the central region, where Gr99\nocite{grillmair99} performed their
analysis, hosts a different population of objects (see the
discussion).

\begin{table}
  \caption{The turnover magnitudes for the complete sample,
    and the blue and red subsamples in $B$, $V$ and $I$.}
  \label{tab.toms}
  \begin{center}
    \begin{tabular}{ccccc}
      \hline
      \noalign{\smallskip}
      Filter & TOM & TOM$_\mathrm{blue}$ & TOM$_\mathrm{red}$ \\
      \hline
      \noalign{\smallskip}
      $B$ & $24.69\pm0.15$ & $24.66\pm0.29$ & $24.82\pm0.31$ \\
      $V$ & $23.87\pm0.20$ & $23.99\pm0.98$ & $23.85\pm0.28$ \\
      $I$ & $22.72\pm0.14$ & $23.16\pm0.47$ & $22.49\pm0.20$ \\
      \hline
    \end{tabular}
  \end{center}
\end{table}

\begin{figure}
  \includegraphics[width=\hsize]{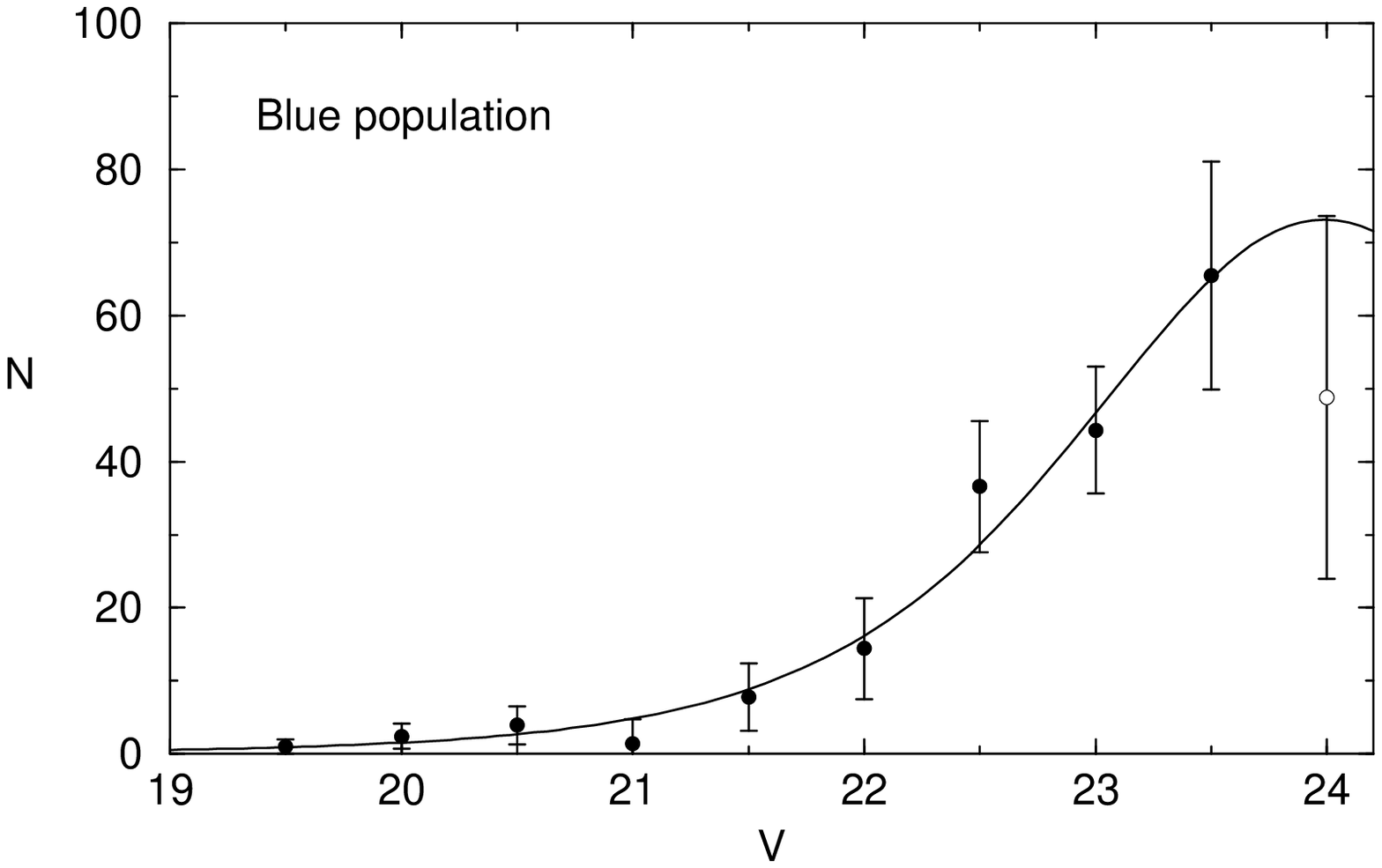}
  \includegraphics[width=\hsize]{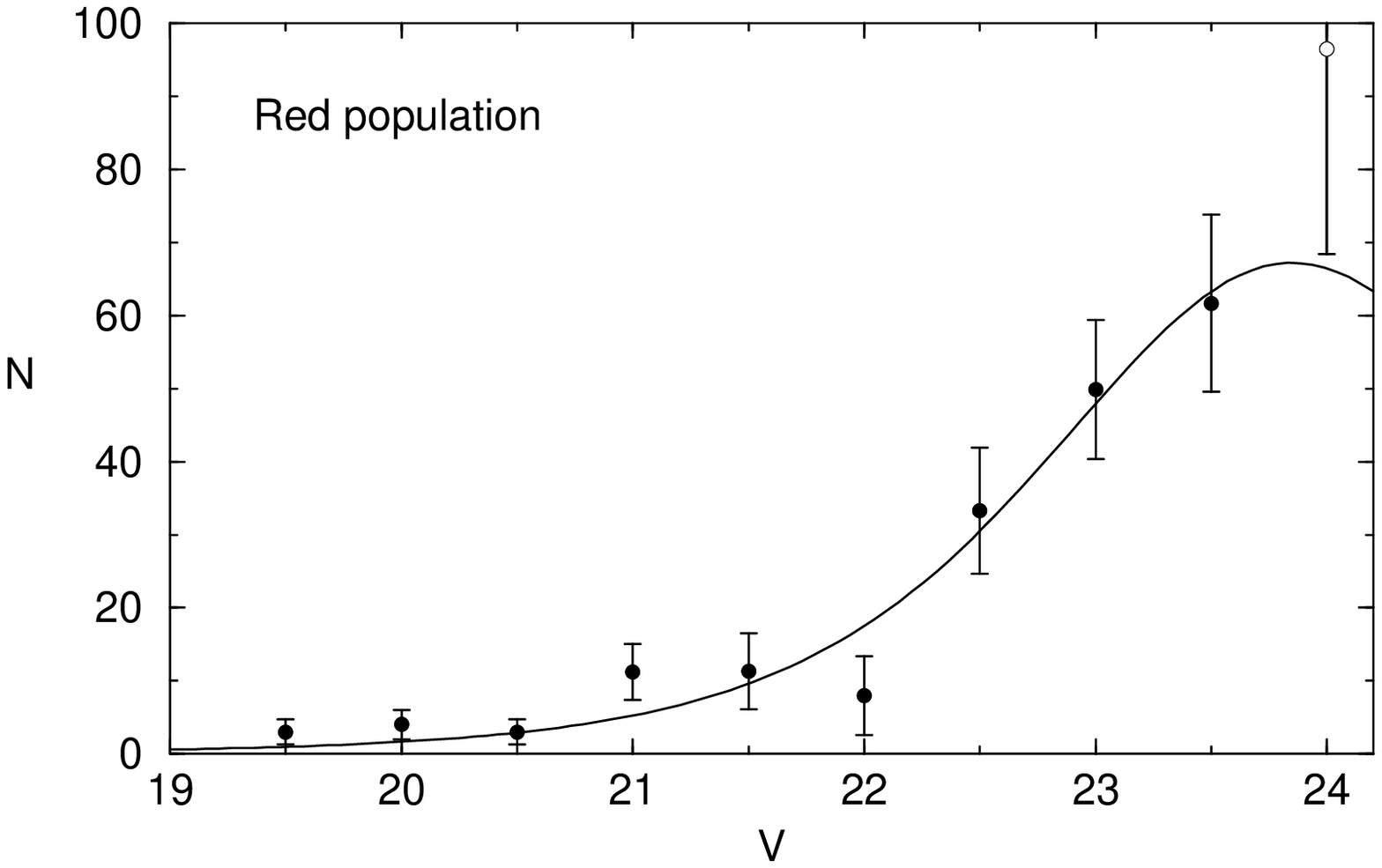}
  \caption{The GCLF of the blue and the red populations,
    in the $V$ band.}
\end{figure}

\begin{table*}
  \caption{Published TOMs for galaxies in the Fornax clusters, using
    the GCLF. References are as following:
    1. Kohle et~al. \cite{kohle96}; 2. Blakeslee \& Tonry
    \cite{blakeslee96}; 3. Elson et~al. \cite{elson98}; 4. Della Valle
    et~al. \cite{dellavalle98}; 5. Ostrov et~al. \cite{ostrov98};
    6. Grillmair et~al. \cite{grillmair99}; 7. Richtler
    et~al. \cite{richtler92}.}
  \label{tab.fornax_toms}
  \begin{center}
    \begin{tabular}{ccccc}
      \hline
      \noalign{\smallskip}
      Galaxy & TOM($B$) & TOM($V$) & TOM($I$) & Reference \\
      \hline
      \noalign{\smallskip}
      \object{NGC\,1344} &                & $23.80\pm0.25$ &                & 2 \\
      \object{NGC\,1379} & $24.95\pm0.30$ & $23.65\pm0.26$ & $22.77\pm0.44$ & 1,3 \\
      \object{NGC\,1380} & $24.38\pm0.09$ & $24.05\pm0.25$ &                & 2,4 \\
                         &                & $23.67\pm0.11$ &                & 4 \\
      \object{NGC\,1399} & $24.59\pm0.10$ & $23.83\pm0.15$ &                & 2,6 \\
                         &                & $23.90\pm0.08$ & $22.27\pm0.05$ & 1 \\
                         &                & $23.71\pm0.12$ &                & 5 \\
      \object{NGC\,1404} & $24.86\pm0.21$ & $24.10\pm0.20$ &                & 6,7 \\
                         &                & $23.92\pm0.20$ &                & 2 \\
      \object{NGC\,1374} &                & $23.44\pm0.13$ & $22.49\pm0.12$ & 1 \\
      \object{NGC\,1427} &                & $23.63\pm0.19$ & $22.22\pm0.13$ & 1 \\ 
      \hline
    \end{tabular}
  \end{center}
\end{table*}

\section{Specific frequency}
\label{sec:specfreq}

The specific frequency {\bf of a GCS} is defined as the number of clusters
per unit luminosity of the parent galaxy referred to $M_V=-15$ (Harris
\& van~den~Bergh \cite{harris81}): $S_N = N\cdot
10^{0.4\cdot(M_V+15)}$, where $N$ is the total number of clusters and
$M_V$ the absolute $V$-magnitude of the host galaxy.

The basis for calculating $N$ is Table~\ref{tab.radialprof}.  We
counted the number of globular clusters in each elliptical ring down
to the TOM in $V$ (see Sects.~\ref{sec:complcorr} and
\ref{sec:lumfunc}).

The data in each ring were then corrected by the corresponding
completeness function. When the rings were not fully covered by our
frames (see Fig.~\ref{isofotas}), a geometrical correction factor was
used to scale the counts to the total area.  The results are listed in
Table~\ref{tab.radialprof}. The raw number of counts are listed in
column~6. Column~7 lists the counts after the completeness correction.
The geometrical factor is given in column~8. To calculate the total
number of clusters in each annulus (column~9), we assumed the
luminosity function to be symmetric around the TOM and doubled our
counts. The total population of clusters (which includes background
objects) was derived as the sum of the counts listed in column~9. The
last three annuli were not included, as the counts cannot be
distinguished from the background. Finally, to estimate the number of
clusters in the central region (first annulus and inner ellipse) we
extrapolated the data to follow the slope of the radial profile
defined by the outer bins. This is because of the poor detection of
clusters in this region, which would make the completeness correction
too uncertain (see Fig.~\ref{completeness}).

Several authors argue that tidal disruption and dynamical effects are
more effective near the centre (for a discussion see e.g. Gnedin
\cite{gnedin97}). Our purpose is to show that even if the central GCs
were not affected by these destruction mechanisms, the specific
frequency is still very low.

We then arrived at a mean density of $160\pm30$\,objects/$\sq\arcmin$
down to the TOM for the innermost bin and inner ellipse. Thus, 
for this region of area $1.407\sq\arcmin$, $225\pm42$ clusters brighter
than the TOM are expected. As with the outer bins, we doubled our
counts around the TOM to obtain $N_{\rm annulus}=450\pm84$.
  
Summing up the results for each annulus gives $1314\pm240$ (see column
9 of Table~\ref{radialprof}).  From this, one only needs to subtract
the background counts, estimated to be $2.0$\,objects/$\sq\arcmin$.
As we considered the geometrical incompleteness of the last three bins
(see Fig.~\ref{isofotas} and column 8 in Table~\ref{radialprof}), the
total area of counting is $35.186\sq\arcmin$ including the nucleus.
Therefore, about $141$ background objects are to be subtracted.
Again, the counts were doubled around the TOM.  Finally, we estimate
$N_T=1173\pm240$ as the total population of clusters in
\object{NGC\,1316}.

With $\mu=31.35$ (Richtler et~al. \cite{richtler00}) and an apparent
magnitude $m_V=8.53$ (de~Vaucouleurs et~al.  \cite{devaucouleur91}),
we obtain $S_N=0.9\pm0.2$, where the uncertainty results from error
propagation and has the form:
\begin{equation}
  \label{eq:uncer}
  \sigma = 10^{0.4\cdot(M_V+15)} \sqrt{\sigma_N^2 
    + (0.4\cdot N \cdot\ln10)^2\cdot\sigma_{M_V}^2}
\end{equation}

This is an extremely low value for a bright early-type galaxy. As we
discuss below, it may result from the fact that the dominant light
contribution to \object{NGC\,1316} comes from an intermediate age
population instead from an old one.  A similar interpretation has
already been made by Hilker \& Kissler-Patig (\cite{hilker96}) in the
case of \object{NGC\,5018}.

Furthermore, if one assumes that, due to dynamical effects, the
surface density of the cluster population does not follow the power
law, but tends to remain approximately constant in the two inner
annuli, then the number of clusters drops to about 150, and the $S_N$
is decreased to 0.7.

Nevertheless, we cannot exclude the possibility of measuring an
erroneous background density. As can be seen from the radial profile
(Fig.~\ref{radialprof}), the background level is reached at $r\sim
300$ arcsec, where the average density is $2.0$\,objects/$\sq\arcmin$.
It is interesting to investigate what are the effects of varying this
number by $1\sigma$, as it enters directly into the Specific Frequency
and hence, into the estimated age of the merger. This is discussed in
Sect.~\ref{sec:discspecfreq}.

\section{Discussion}
\label{sec:disc}

\subsection{Subpopulations of clusters}

The detection of intermediate-age clusters by Go00\nocite{goud00}
clearly shows that the cluster system is inhomogeneous in terms of
age. Also the strikingly different angular distributions of the blue
and red clusters argues in favour of the existence of subpopulations.
One would naturally assume that the red and probably metal-rich
distribution, or part of it, has been formed in the merger. A popular
scenario attributes the observed bimodal colour distributions in many
early-type galaxies as being due to cluster formation in enriched gas
during a merger (Ashman \& Zepf \cite{ashmanzepf92},
\cite{ashmanzepf97}). This bimodality is not seen in
\object{NGC\,1316}.  However, we think that this is not of great
significance, given the following points.

A bimodal metallicity distribution as a result of cluster formation
during a merger probably needs special conditions. If, for example,
the merger components were two spiral galaxies with a certain range of
metallicities present in the gas content of their disks, then one
naturally expects cluster formation with a similar range of
metallicities, perhaps weighted to higher metallicities due to
chemical enrichment during the merger. Any bimodal metallicity
distribution, which has been present before the merger (and which does
not stand in relation with the merger, like the bulge and disk
clusters in the Milky Way), could well have been blurred.

Moreover, a metal-rich intermediate-age population of clusters, formed
in the merger, for example with an age of 2\,Gyr, would have $B-V$
colours which are about $0.2$\, mag bluer than their old, metal-rich
counterparts. According to the above relation between $B-I$ colour and
[\element{Fe}/\element{H}], this corresponds to a metallicity
difference of $0.5$\,dex.  The peak at $-0.8$\,dex in
Fig.~\ref{metal_histo} could well be explained in this way, comprised
perhaps by clusters with solar metallicity and those found by
Go00\nocite{goud00}.

\subsection{The Specific Frequency}

\label{sec:discspecfreq}

Our result of $0.9\pm0.2$ for the specific frequency of the GCLF of
\object{NGC\,1316} confirms with a larger field what has been found by
Gr99\nocite{grillmair99} using HST-data. This is an extremely low
value for a bright, early-type galaxy, even more astonishing as one
would expect to have many clusters formed during the merger process.

One possible explanation for this low value would be to assume that
the merger caused, whether or not star clusters have been formed, a
period of star formation, which nowadays constitute an
intermediate-age population, which is brighter than an old population
without strikingly differing in integral broad-band colours. The
contribution of such a population could be visible spectroscopically
by stronger hydrogen lines than normal.

We assume that a ``normal'' specific frequency of early-type galaxies
is 4.  The question then is: How young must \object{NGC\,1316} be to
lower that value to our value of 0.9?  Worthey (\cite{worthey94})
offers a tool for synthesizing stellar populations. Assuming a
metallicity of $+0.5$\,dex, we get from his ``model interpolation
engine'' an age of 2\,Gyr to account for the magnitude difference of
$1.6$\,mag in comparison to an old population of 15\,Gyr.

The dominant error in this result is the uncertainty in the background
density. If it were, instead of $2.0$\,objects/$\sq\arcmin$,
$4.0$\,objects/$\sq\arcmin$, the total number of clusters decreases to
$N_{T}=1033$ and $S_{N}=0.8$.  On the other hand, a density of
$0.0$\,objects/$\sq\arcmin$ implies $N_{T}=1314$ and $S_{N}=1.0$.
Thus, a total error of $\sim 0.3$ results for $S_{N}$. This means,
from Worthey's models, an error in the age of the merger of
$\sim$0.8\,Gyr.

Our value of 2\,Gyr is in good agreement with the luminosity-weighted
mean stellar age of \object{NGC\,1316} given by Kuntschner \& Davies
(\cite{kuntschner98}), who measured line strengths for early-type
galaxies in the Fornax cluster.  It also matches the merger age of
\object{NGC\,1316} estimated by Mackie \& Fabbiano (\cite{mackie98})
and the globular cluster ages of Go00\nocite{goud00}.  In this
context, one also should note that the occurence of two SNe Ia is a
further indication for the existence of an intermediate-age
population, to which many SNe Ia seem to belong (e.g. McMillan \&
Ciardullo \cite{mcmillan96}).

The question now arises whether specific frequencies of GCS in general
could be influenced, if not dominated, by the existence of
intermediate-age populations. One notes that among the sample of
early-type galaxies, for which Trager et~al. (\cite{trager2000})
measured age-sensitive line strengths, a few striking examples like
\object{NGC\,221} or \object{NGC\,720} exist, which have a young
component and whose specific frequencies are quite low for early-type
galaxies. Another example is the merger candidate \object{NGC\,5018},
for which Hilker \& Kissler-Patig (\cite{hilker96}) found a low
specific frequency, similar to that of \object{NGC\,1316}, which they
also explained by the presence of a young population.

A strong argument against the conjecture that this effect is
dominating, is the fact that the specific frequency among early-type
galaxies has a positive correlation with the environmental galaxy
density (West \cite{west93}).  If an enhanced number of mergers in a
high density environment caused the existence of intermediate-age
populations in many early-type galaxies, one would expect a negative
correlation, if the brightness of the host galaxy significantly
influenced the specific frequency.  However, if the efficiency of
globular cluster formation increases with the star formation rate, as
the work of Larsen \& Richtler (\cite{soeren00}) indicates, then a
positive correlation, as observed, is understandable.

Larsen \& Richtler (\cite{soeren00}) performed a survey of young
clusters in a sample of 21 nearby face-on spiral galaxies. They found
that the ``specific luminosity'' $T_L(U)$, which measures the ratio of
the summed-up U-luminosity of cluster candidates to the total
U-luminosity of the host galaxy, correlates well with indices
measuring the star formation rate.  Apparently, a high star formation
rate causes also a higher cluster formation efficiency.

The final specific frequency after a merger therefore depends on many
circumstances, among them the star formation rate, the amount of gas
available for star formation, the specific frequency of the merger
components, etc.

Can we expect that the merger enhances the specific frequency of the
outcoming configuration?  Go00\nocite{goud00} describe the evidence
that makes \object{NGC\,1316} a candidate for an advanced disk-disk
merger system. So let us assume that two spiral galaxies of equal
brightness $M_V$ have merged.

If we set for the visual luminosity $L = 10^{-0.4 \cdot M_V}$ then the
following relation holds:
\begin{equation}
  N_\mathrm{tot} - N_\mathrm{m} = 2 \cdot 10^{-6} \cdot S_{N} \cdot L
\end{equation}
with $N_\mathrm{tot}$ the total number of clusters, $N_\mathrm{m}$ the
number of cluster formed in the merger, and $S_{N}$ the specific
frequency of the merging spiral galaxies. Further:
\begin{equation}
  L_\mathrm{tot} = 2 \cdot L_\mathrm{dim} + L_\mathrm{m}
\end{equation}
where $L_\mathrm{tot}$ is the actual luminosity of \object{NGC\,1316},
$L_\mathrm{dim}$ the actual luminosity of the former spirals, dimmed
for about 3\,Gyr, after star formation ceased, and $L_\mathrm{m}$ is
the luminosity of the stellar population formed in the merger. A
typical value for $S_{N}$ is 0.5 (Harris \cite{harris00}).
$N_\mathrm{m}$ is not accurately known, but a rough estimate comes
from the fact that Go00\nocite{goud00} found 3 clusters of masses
comparable with $\omega$ Cen, so a lower boundary is 300 clusters.
With these numbers, the luminosity of the hypothetical spirals, at the
time of merging, turns out as $9\cdot 10^{8}$, corresponding to
$M_{V}=-22.4$. The dim factor during the next 3\,Gyr is about 2, based
on a simulation with Worthey's ``model interpolation engine''. If
$L_\mathrm{tot}$ is $1.32\cdot 10^{9}$, then $L_\mathrm{m}$ turns out
as $4.2 \cdot 10^{8}$, corresponding to $M_V=-21.6$. The stellar
population formed in the merger will dim for 1.5 mag during the next
12\,Gyr. The further fading of the spiral will be roughly by factor of
5, corresponding to the ratio of 10 between the M/L values of a spiral
and an elliptical galaxy. Thus the total luminosity after aging will
be about $3 \cdot 10^8$, which means $M_V=-21.2$ and a specific
frequency of 4.

The numbers used in this exercise are uncertain, but it shows that the
higher specific frequency of elliptical galaxies with respect to
spirals can indeed plausibly be understood as the outcome from
disk-disk mergers.

The example also shows that a merging scenario of an elliptical galaxy
with a spiral leads to such high merger luminosities, that the gas
mass consumed by the star formation in the merger cannot reasonably
be brought in by a spiral galaxy. For instance, an elliptical
with $M_{V} = -21.5$ and specific frequency 3 corresponds through the
above relations to a spiral galaxy with $M_{V} = -21.2$ and a merger
brightness of $M_{V} = -22.4$. A population of this brightness with an
age of 3\,Gyr has a mass of almost $8 \cdot10^{10}\,M_\odot$, which
cannot be provided by any ordinary-sized spiral.

It is, however, doubtful, whether high specific frequencies (6 and
more) can be created by the merging of two or more spiral galaxies
(Harris \cite{harris00}).  It is more likely that dwarf galaxies must
be involved if cluster systems like those of \object{NGC\,1399} and
\object{M87} are to be formed (Hilker et~al. \cite{hilker99}).

\subsection{The Luminosity Function}

The importance of the LF of \object{NGC\,1316} lies in the fact that
it could provide an accurate distance to the two SNe Ia
\object{SN\,1980N} and \object{SN\,1981D} and thus help to calibrate
the absolute SN Ia luminosity, as has been the case for
\object{SN\,1992A} in \object{NGC\,1380} (Della Valle et~al.
\cite{dellavalle98}).

However, due to its presumed merger history, one might suspect that
the GCS of \object{NGC\,1316} is contaminated by younger clusters,
which very probably should be searched for among the red cluster
population. Broad band colours are not very sensitive: a cluster
becomes redder in $V-I$ by only $0.2$\,mag, evolving from an age of
2\,Gyr to an age of 15\,Gyr.

Such a contamination would also affect the turnover magnitude.
OT95\nocite{okazaki95} investigated the evolution of a GCS considering
the dynamical and photometric evolution of individual clusters.
According to their results, the TOM would brighten by about
$0.6$--$1.0$\,mag, when a GCS evolves from an age of 2\,Gyr to an age
of 15\,Gyr. Therefore one expects the TOM of the red cluster
population to be fainter than that of the blue population, if the red
clusters are indeed younger.

This has to be proved/disproved with deeper data than those at our
disposal. Since the TOM is not reached in either among the red
population or the blue population, its numerical value returned by the
fit is rather uncertain and hence we cannot make any secure statement.

From the models of OT95\nocite{okazaki95}, we would expect to see the
TOM approximately at $V=24.5$ or $B=25.5$ in case that the majority of
the red clusters has been formed in the merger.

Gr99\nocite{grillmair99}, in a small field near the centre of
\object{NGC\,1316}, obtained $B$-photometry with HST to faint
magnitudes. Their results are not easy to interpret. Their actual
background-corrected and completeness-corrected counts do not 
rule out a turnover at $B=25.5$, but no turnover is indicated 
in their $I$-counts at $I=23.7$, where one expects it to be due to 
the mean $B-I$ colour of 1.8. If they correct for the contribution of
``dust-obscured'' objects, the resulting LF is completely dominated by
the unseen population and there is a steady increase in both $B$ and
$I$ down to the faintest magnitudes.

Gr99\nocite{grillmair99} conjecture that they see a population of open
clusters, which is plausible if the star formation near the center has
continued for some time after the merger. One can imagine that most of
the globular clusters formed during the merger, when the star
formation rate was high, but less massive open clusters continued to
form at lower rate, which would blur the luminosity function of the
globular clusters.

No detailed picture follows from the present data, but it is not
excluded that a large fraction of the red globular clusters formed
during the merger. Another (unknown) part may have been brought in by
the merger components. If two spiral galaxies have merged they could
have brought in their metal-rich bulge populations. The blue globular
clusters would then be the common halo globular clusters of the two
merger components. Star formation near the centre continued and
younger star clusters were added to the cluster population formed
during the merger. In that case, we would expect the outer population
of red clusters to exhibit a TOM in the range 0.6 to 1\,mag fainter
than the blue cluster population, which could be shown by deeper
photometry than we have.

In this scenario one should  only regard the LF of the blue cluster
population as appropriate for distance estimation.

\subsection{A remark on the supernovae in \object{NGC\,1316}}

Two SNe of type Ia, \object{SN\,1980N} and \object{SN\,1981D},
appeared in \object{NGC\,1316} (Hamuy et~al. \cite{hamuy91}). The
distance determined by its GCLF can thus contribute to the calibration
of the absolute peak brightness of SN Ia. There are only a handful of
nearby early-type host galaxies for Ia supernovae, so
\object{NGC\,1316} occupies an important role.

However, our data are not deep enough to measure the distance via the
GCLF so accurately that \object{SN\,1980N} could be independently
compared with \object{SN\,1992A} in \object{NGC\,1380}, the third SN
Ia which appeared in the Fornax cluster, and which has a very well
measured GCLF (Della~Valle et~al. \cite{dellavalle98}).
\object{SN\,1981D} has a light curve of lower quality.

As Tables~\ref{tab.toms} and \ref{tab.fornax_toms} show, the TOM of 
the GCS of \object{NGC\,1316} fits at least well to the other TOMs 
of early-type Fornax galaxies. This may be taken as an argument that
\object{NGC\,1316}, in spite of its outlying location, is in the same
distance as the core of the Fornax cluster.

Indeed, after the appropriate corrections for decline-rate and colour
(e.g. Hamuy et~al. \cite{hamuy96}, Richtler et~al. \cite{richtler00}),
the apparent peak brightness of \object{SN\,1980N} agrees excellently
with that of \object{SN\,1992A}, which is a much stronger argument for
their equal distance.

\begin{acknowledgements}
We thank the referee Alfred Rosenberg for numerous comments 
that helped to improve the clarity of this article. 
Russell Smith is gratefully acknowledged for a careful reading of the 
manuscript. M.G. thanks the Sternwarte der Universit\"at 
Bonn for its hospitality, as well as the 
Deutscher Akademischer Austauschdienst (DAAD) for a studentship. 
L.I. thanks FONDECYT for support through "Proyecto Fondecyt 1890009".
\end{acknowledgements}

\end{document}